\documentclass{aastex6}

\begin{document}

\title{A New Measurement of the Spectral Lag of Gamma-Ray Bursts and its Implications for Spectral Evolution Behaviors }

\author{\sc Lang Shao\altaffilmark{1,2}, Bin-Bin Zhang\altaffilmark{3,4}, Fu-Ri Wang\altaffilmark{1}, Xue-Feng Wu\altaffilmark{2,5,6}, Ye-Hao Cheng\altaffilmark{1}, Xi Zhang\altaffilmark{1}, Bang-Yao Yu\altaffilmark{1}, Bao-Jia Xi\altaffilmark{1}, Xue Wang\altaffilmark{1}, Huan-Xue Feng\altaffilmark{1,7}, Meng Zhang\altaffilmark{1}, Dong Xu\altaffilmark{7}}
\altaffiltext{1}{Department of Space Sciences and Astronomy, Hebei Normal University, Shijiazhuang 050024, China}
\email{lshao@hebtu.edu.cn (L.S.)}
\altaffiltext{2}{Purple Mountain Observatory, Chinese Academy of Sciences, Nanjing 210008, China}
\altaffiltext{3}{Instituto de Astrof\'isica de Andaluc\'a (IAA-CSIC), P.O. Box 03004, E-18080 Granada, Spain}
\altaffiltext{4}{Scientist Support LLC, Madsion, AL 35758, USA}
\altaffiltext{5}{School of Astronomy and Space Sciences, University of Science and Technology of China, Hefei 230026, China}
\altaffiltext{6}{Joint Center for Particle, Nuclear Physics and Cosmology, Nanjing University-Purple Mountain Observatory, Nanjing 210008, China}
\altaffiltext{7}{CAS Key Laboratory of Space Astronomy and Technology, National Astronomical Observatories, Chinese Academy of Sciences, Beijing 100012, China}

\begin{abstract}

We carry out a systematical study of the spectral lag properties of 50 single-pulsed Gamma-Ray Bursts (GRBs) detected by Fermi/GBM. By dividing the light curves into multiple consecutive energy channels we provide a new measurement of the spectral lag which is independent on energy channel selections. We perform a detailed statistical study of our new measurements. We find two similar power-law energy dependencies of both the pulse arrival time and pulse width. Our new results on the power-law indices would favor the relativistic geometric effects for the origin of spectral lag. However, a complete theoretical framework that can fully account for the diverse energy dependencies of both arrival time and pulse width revealed in this work is still missing. We also study the spectral evolution behaviors of the GRB pulses. We find that the GRB pulse with negligible spectral lag would usually have a shorter pulse duration and would appear to have a ``hardness-intensity tracking'' (HIT) behavior and the GRB pulse with a significant spectral lag would usually have a longer pulse duration and would appear to have a ``hard-to-soft'' (HTS) behavior.

\end{abstract}


\section{Introduction} \label{sec:intro}

Gamma-ray bursts (GRBs) are prolific emitters of non-thermal radiation found in cosmological distance that can span for many decades in frequency. The prompt emission phase of GRBs is still not well understood given their complex temporal structures and spectral properties \citep[see][for recent reviews]{2015PhR...561....1K,2015AdAst2015E..22P}. Study of the prompt GRB spectra is essential for understanding the radiation mechanisms and their physical origins. In general, the integrated spectrum could be well described by the so-called Band function, which is a smoothly joint broken power law peaking at $E_{\rm p}$ in the $\nu f_{\nu}$ spectrum \citep{1993ApJ...413..281B}. This peak energy $E_{\rm p}$ is usually well correlated with the estimated isotropic total GRB energy ($E_{\rm \gamma,iso}$), and peak luminosity ($L_{\rm p}$) \citep{2002A&A...390...81A,2003MNRAS.345..743W,2004ApJ...609..935Y,2006MNRAS.372..233A,2012MNRAS.421.1256N,2012ApJ...750...88Z}. 

As GRBs are typically highly variable in intensity and significant spectral evolution apparently prevails, time-resolved spectral analyses have been considered more crucial to crack the radiation mechanism of GRBs. The correlation between the time-resolved peak energy and the corresponding isotropic peak luminosity could be well-established within individual bursts \citep[e.g.,][]{2010A&A...511A..43G,2012ApJ...754..138F,2015ApJ...807..148G}. Multiple spectral components have been proposed by time-resolved analysis of data from different satellites/detectors \citep[e.g.,][]{2011ApJ...730..141Z,2013A&A...550A.102T,2013ApJ...779..175F,2014MNRAS.442..419B,2015MNRAS.450.1651I,2015ApJ...814...10G,2016A&A...588A.135Y}. Different radiation mechanisms have been tried to interpret the spectral shape \citep[e.g.,][]{2011MNRAS.415.3693R,2012ApJ...758L..34Z,2013ApJ...770...32G,2013ApJ...765..103L,2014NatPh..10..351U,2014ApJ...784...17B,2015A&A...573A..81Y,2015ApJ...813..127P,2015ApJ...802..132C,2016ApJ...816...72Z,2016ApJ...819...79G}. 

As abundant information about the spectral properties has been revealed by time-resolved analysis, complexity and unexpected confusion have also been introduced. Different analysis methods including using different time-binning schemes, samples, energy ranges can lead to different results. The shape of the spectra and/or the pattern of the spectral evolution have been found to be affected by the blend of different pulses. Two different correlations between the spectral peak energy and the corresponding luminosity have been proposed:``hard-to-soft'' (HTS) and ``hardness-intensity tracking'' (HIT), which are seemingly incompatible with each other \citep{2011ApJ...740..104H,2012ApJ...756..112L,2015ApJ...815..134H,2015MNRAS.447.3087B}. Binning method is also found to play an important role in analyzing the pattern of the spectral evolution \citep{2014MNRAS.445.2589B}. 

Since the temporal profiles of GRBs are very complicated and the spectral property is difficult to identify in blended pulses, efforts have been made in analyzing the bursts with a single or well-separated pulse. Both the pulse width and the arrival time have been found dependent on energy in most GRBs, revealing an important connection between the temporal profile and the spectral evolution \citep{1995A&A...300..746C,1995ApJ...448L.101F, 1996ApJ...459..393N,1997ApJ...486..928B}. Spectral lag can be measured between two given energy ranges by cross-correlating the light curves of two corresponding energy channels. Spectral lag has been found well correlated with the peak luminosity \citep{2000ApJ...534..248N,2002ApJ...579..386N,2010ApJ...711.1073U}. The spectral lag has also been considered as an important tool to classify long and short GRBs \citep{2006Natur.444.1044G,2006ApJ...643..266N,2006MNRAS.373..729Z,2015MNRAS.446.1129B}. Detailed analysis of the spectral lag would utilize both the temporal and spectral information and help to reveal the radiation mechanisms of GRBs. The existence of the spectral lag can be considered as a consequence of the spectral evolution in the radiation processes \citep[e.g.,][]{1998ApJ...501L.157D,2003ApJ...594..385K,2003MNRAS.342..587D,2005A&A...429..869R,2016ApJ...825...97U,2016A&A...592A..95M}. For synchrotron and synchrotron self-Compton (SSC) processes in relativistic plasma outflows, the pulse width should be correlated with the photon energy $E$ as $E^{-1/2}$ and $E^{-1/4}$ , respectively \citep{1998ApJ...493..708K,1998ApJ...508..752C,1998ApJ...501L.157D}. On the other hand, spectral lags can also be explained by some geometric effect including the curvature effect \citep[e.g.,][]{2005MNRAS.362...59S,2006MNRAS.367..275L,2004ApJ...617..439Q} or the pulse confusion of different spectral components \citep{2013ApJ...770...32G,2008ApJ...689L..85E}.

In the previous works, the spectral lag is calculated using Cross-Correlation Function (CCF) between two given energy channels. The value of the spectral lag is highly sensitive to the energy channels which are chosen arbitrarily and a GRB would have a different lag if different energy channels are selected. To take into account such effect and further reveal the observational details on the spectral evolution, we carry out a systematic research of 50 bright GRB pulses detected by Fermi Gamma-Ray Burst Monitor (GBM) and perform a new analysis of the spectral lag using  the light curves in nine consecutive energy channels for each GRB.  By investigating the energy dependencies of both the arrival time and pulse width in universal forms, we are able to provide a new description of the spectral lag effect over all energy channels for each burst independent on energy channel selections. Details of the sample selection and data reduction are presented in Section 2, followed by the description of our analyses in Section 3. We discuss the implications of the spectral hardness evolution behaviors in Section 4. Brief discussion and conclusion are summarized in \S 5.

\section{Sample Selection and Data Reduction} \label{sec:sample}

\begin{figure}
\figurenum{1}
\fig{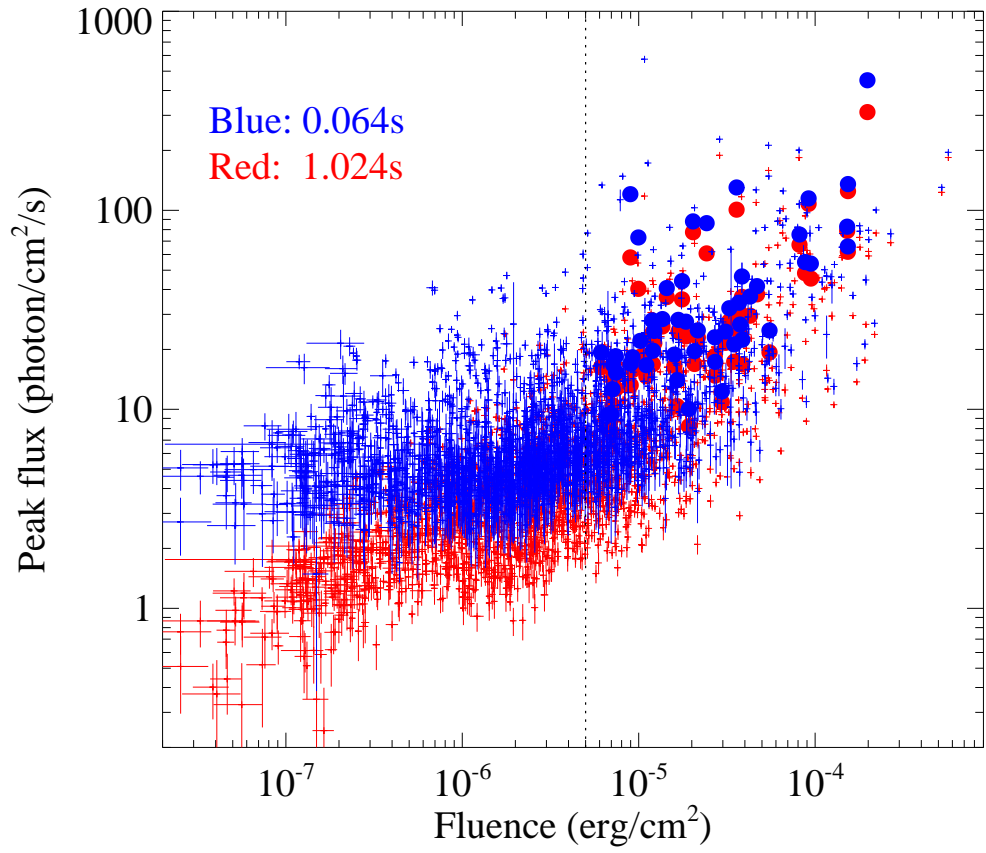}{0.8\textwidth}{}
\caption{Correlation between the total fluence and the peak fluxes in two different time resolutions (0.064~s in blue and 1.024~s in red). The 50 bright bursts in our sample as listed in Table~\ref{tab:sample} are marked with filled circles. The other 1917 bursts detected by Fermi/GBM as of 18 December 2016 are marked with plus signs. The vertical dotted line indicates our selection by the total fluence with $F>5\times 10^{-6}\,{\rm erg\,cm^{-2}}$.\label{fig:fluxfluence}}
\end{figure}

This work made an extensive use of the data from the Gamma-ray Monitor (GBM) on board Fermi Gamma-ray Space Telescope \citep{2009ApJ...702..791M}. For the first step, we searched in the official GBM online burst catalog \citep{2014ApJS..211...12G,2014ApJS..211...13V} for bright bursts with a total fluence $F>5\times 10^{-6}\,{\rm erg\,cm^{-2}}$ in 10-1000 keV. For simplicity, we selected the bright bursts based on their total fluence instead of their peak flux, since the total fluence is generally a well determined property. On the other hand, the peak flux is significantly dependent on the selection of time resolution and therefore has several different values in different time intervals for each burst as provided in the burst catalog. As shown in Figure~\ref{fig:fluxfluence}, the total fluence is generally well correlated with the peak fluxes in different time resolutions. Our sample (as will be introduced below) is a bright sample based on both measures. For the second step, we then carefully selected the ones with only one well-defined pulse by manually examining their light curves recorded in their brightest sodium iodide [NaI(Tl)] detector. For each burst, we only used the time-tagged event (TTE) data of the brightest NaI detector. The TTE data consist of individual photon arrival times with 2~${\rm \mu s}$ temporal resolution and 128-channel spectral resolution, recorded from about 30~s before to 300~s after the burst trigger. 

By selecting the most bright bursts based on their total fluence, we have included in our sample only one short ($<2$~s) burst - GRB 140209313 with a duration of $\sim$~1.4 s and a fluence of $\sim\,9\times 10^{-6}\,{\rm erg\,cm^{-2}}$, which turns out to be the most bright short burst (without considering the distance) detected in GBM as of 30 May 2016. There are six short bursts in the online burst catalog that meet the fluence criterion. Among them, GRBs 090227772 ($T_{90}\sim 1.3$~s), 090228204 ($T_{90}\sim 0.4$~s), 120624309 ($T_{90} \sim0.6$~s), 140901821 ($T_{90}\sim 0.2$~s) and 150819440 ($T_{90}\sim 1.0$~s) all have multiple pulses. We have excluded these bursts without carrying out further analyses. The fluence of previously well-studied short burst 090510016 is only $\sim\,3\times 10^{-6}\,{\rm erg\,cm^{-2}}$ and also has multiple overlapping pulses, therefore does not meet our the criterion. Most of other long bursts have also been excluded due to the same reason.

As the third and the most important step for our sample selection, the TTE data for each burst were re-binned into nine energy channels evenly separated in the logarithmic scale. The light curve in each channel was then fitted independently by a first-order polynomial (to subtract the background) together with a Gaussian function $f(t)$ (to pinpoint the pulse):
\begin{equation}
f(t) = a\,e^{-{(t-t_{\rm peak})^2 \over 2\delta^2 }},
\end{equation}
where $a$ is a constant and $t_{\rm peak}$ and $\delta$ represent the arrival time of the peak of the pulse and the RMS width of the Gaussian shape, respectively. 

In generally, the typical shape of a single GRB pulse was proposed to be asymmetric, e.g., with a fast rise and an exponential decay as in the BATSE GRBs \citep{1994ApJS...92..229F}. A two-sided exponential or Gaussian profile with the decay time significantly larger than the rise time (with a ratio ranging from about 2 to 3) was also proposed to be suitable for BATSE GRBs \citep[e.g.,][]{1996ApJ...459..393N, 2003ApJ...596..389K, 2005ApJ...627..324N}. During our fitting processes, we have tried several asymmetric functions to  fit the pulse profiles, but none of them worked perfectly for most the lightcurves in our sample. Indeed, we found that a Gaussian profile was adequate for pinpointing the bulk photons of the pulses with utilizing the standard IDL fitting function, \texttt{GAUSSFIT}, and a more complex function was not necessary. Unfortunately, many very bright pulses that previously appeared to be single as manually selected from the whole TTE data with a very raw time resolution in the second step would still fail to be pinpointed by a Gaussian function in one of the nine energy channels as the time binsize was lowered in the third step. Pulses would also become very noisy and undistinguishable from the background when the binsize is very small. As a standard, the binsize in our work has been automatically set as $5\,\%$ of the duration $T_{90}$ for each burst. The $T_{90}$ is the duration of the time interval during which the detector accumulates from the $5\%$ to the $95\%$ of the photons in a given energy range (50-300 keV for GBM), which has been provided by the official GBM online burst catalogs \citep{2014ApJS..211...12G,2014ApJS..211...13V}. Since the $T_{90}$ is not always representative of the width of the single pulse, some modifications to the bin sizes have been made manually to keep them close to $5\,\%$ of the genuine pulse width. Given a determined time bin, any burst that has one or more pulses that could not be well pinpointed by the Gaussian profile was excluded as a final step of the sample selection process. 

Since each burst has a different range of spectral distribution, the most upper and lower edges of the energy channels were selected manually to ensure enough signal above background (S/N $>$ 5) in each channel. Table~\ref{tab:sample} lists our final sample of 50 bursts (between August 2008 and May 2016) that are qualified for our selection criteria above. The final binsize of each burst is listed in the third column in Table~\ref{tab:sample}. The nine-channel light curves of each burst are shown in the left panels in Figure Set~\ref{fig:sample} where the total light curve summed in the nine channels is shown on the top. The Gaussian profile fitting is shown on top of each light curve in red curve. For a visual aid to manually check the effect of spectral lag, the peaks of each Gaussian profile are connected and plotted in green dashed lines.

\section{New Measurement of the Spectral Lag} \label{sec:style}
As the pulses in each energy channel have been well fitted by Gaussian profiles for each burst, the peak arrival time $t_{\rm peak}$ and the RMS Gaussian width $\delta$ could be well determined. Both $t_{\rm peak}$ and $\delta$, each as a function of the photon energy $E$ (the mid-value of each energy channel), are shown in the right panels of Figure Set~\ref{fig:sample} for each burst. The uncertainties in $t_{\rm peak}$ and $\delta$ are the 1-$\sigma$ errors evaluated for the returned parameters from the Guassian fitting using the standard IDL fitting function {\it GAUSSFIT}. An anti-correlation between $t_{\rm peak}$ and $E$ apparently exists for almost all the bursts. We fitted this anti-correlation with a three-parameter power-law, 
\begin{equation}
t_{\rm peak}(E) = t_0+\tau\times ({E\over 1\,{\rm keV}})^{-\beta},\label{eq:powerlaw1}
\end{equation}
where, $t_0$ and $\tau$ are constants in the unit of second and $\beta$ is the power-law index.
We notice that there are some practical meanings of $t_0$ and $\tau$. If $\beta>0$, then we have
\begin{equation}
t_{\rm peak}\,({\rm 1 keV}) = t_0 + \tau,\label{eq:t1keV}
\end{equation}
and
\begin{equation}
t_{\rm peak}(\infty) = t_0, \label{eq:tinf}
\end{equation}
in two limiting cases with the photon energy of 1~keV and $\infty$, respectively. So $t_0\tbond t_{\rm peak}(E=\infty)$ can be regarded as the limiting value of the earliest arrival time of the most energetic photons. $\tau$ is the difference between $t_{\rm peak}(1\,{\rm keV})$ and $t_{\rm peak}(\infty)$, thus it is the ``limiting'' spectral lag at 1~keV. We use $\tau$ as a fundamental timing measurement of the spectral lags independent on the energy channel selections.

There is also an apparent anti-correlation between $\delta$ and $E$ for almost all the bursts in our sample. We fitted this anti-correlation with a two-parameter power-law,
\begin{equation}
\delta (E) = \omega \times ({E\over 1\,{\rm keV}})^{-\gamma},\label{eq:powerlaw2}
\end{equation}
where, $\omega$ is a constant in the unit of second and $\gamma$ is the power-law index. $\omega$ could be considered as the limiting half pulse width at 1~keV.

\begin{figure}
\figurenum{3}
\gridline{\fig{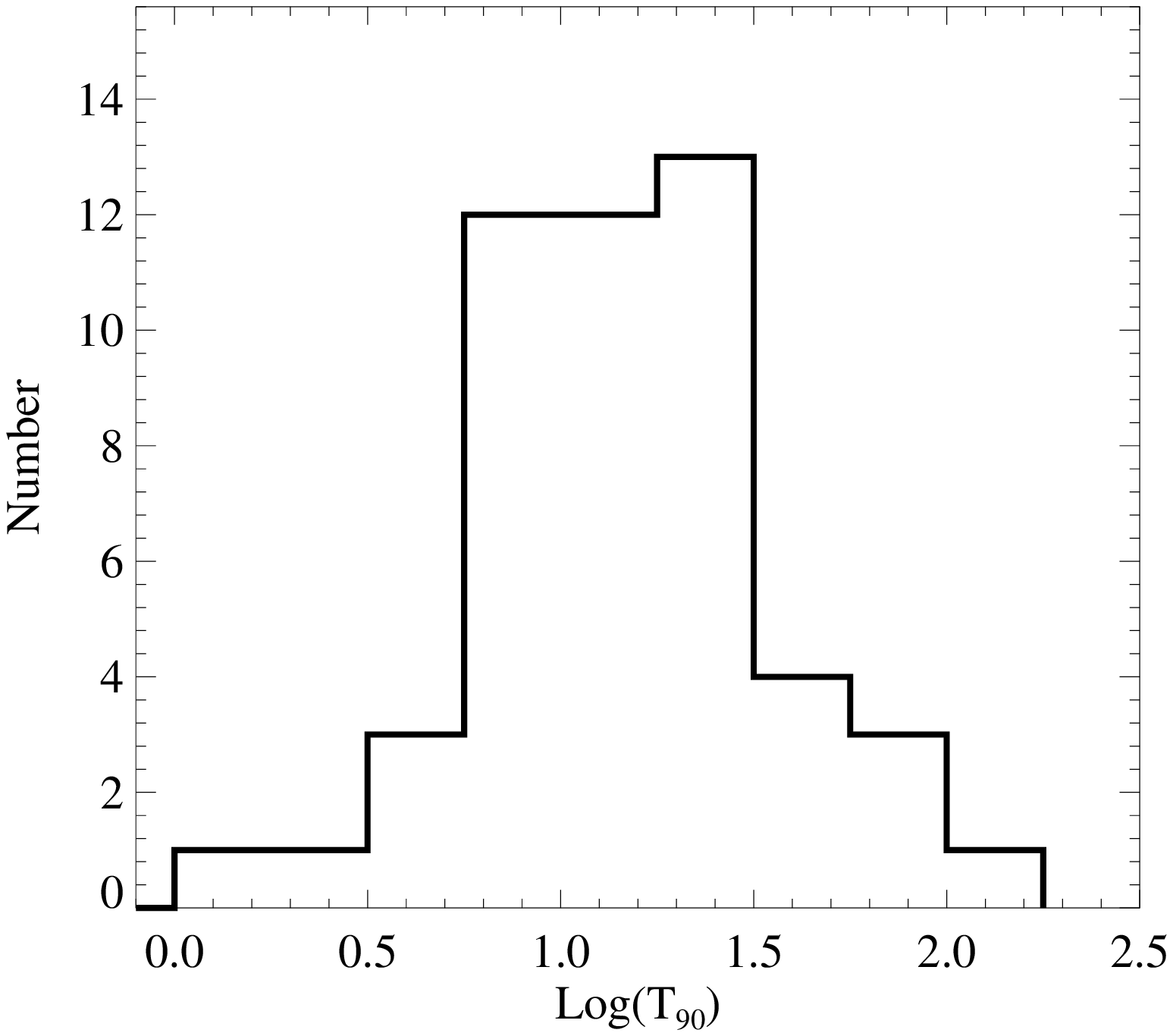}{0.45\textwidth}{(a)}
			\fig{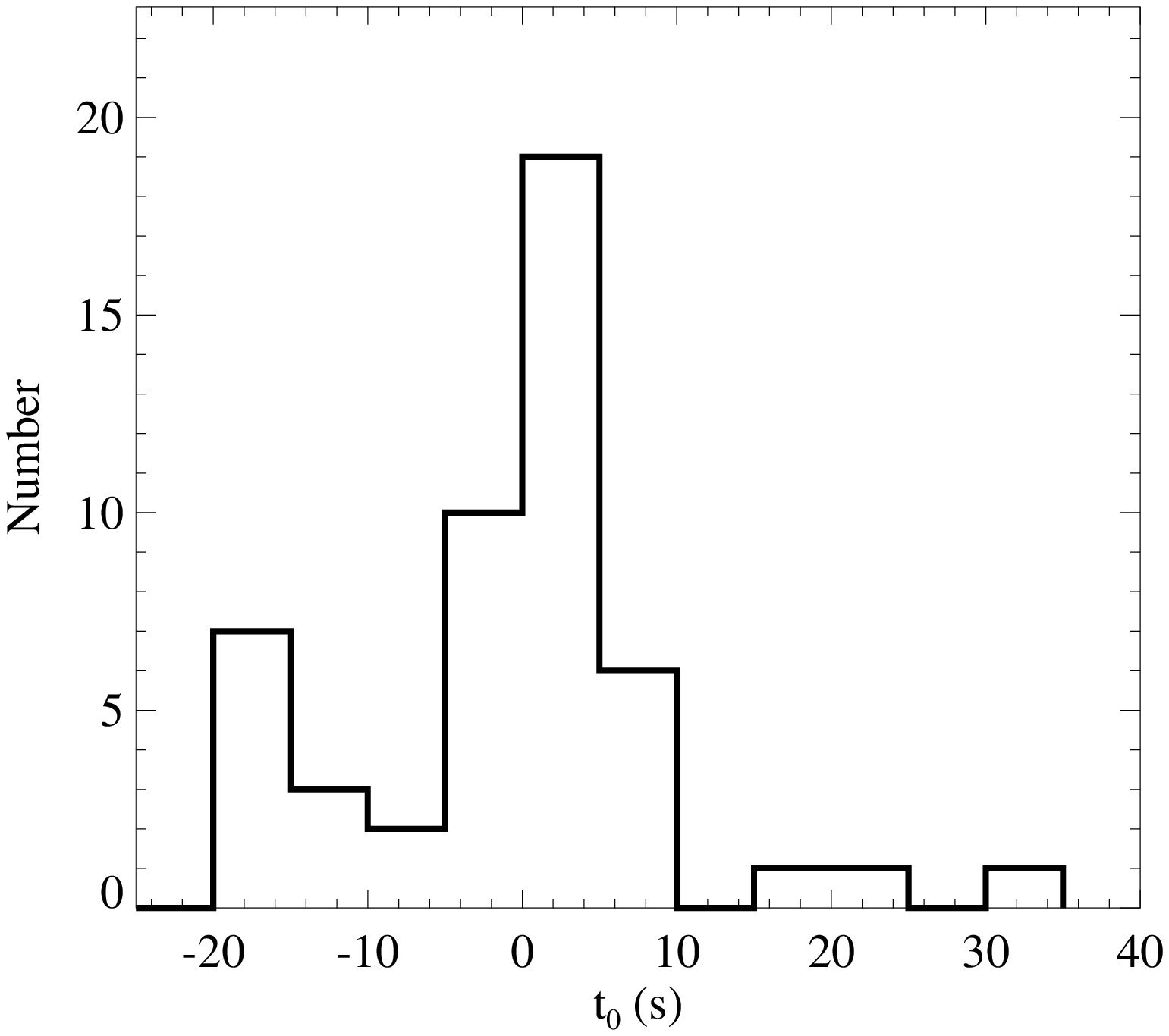}{0.45\textwidth}{(b)}
        }
\gridline{\fig{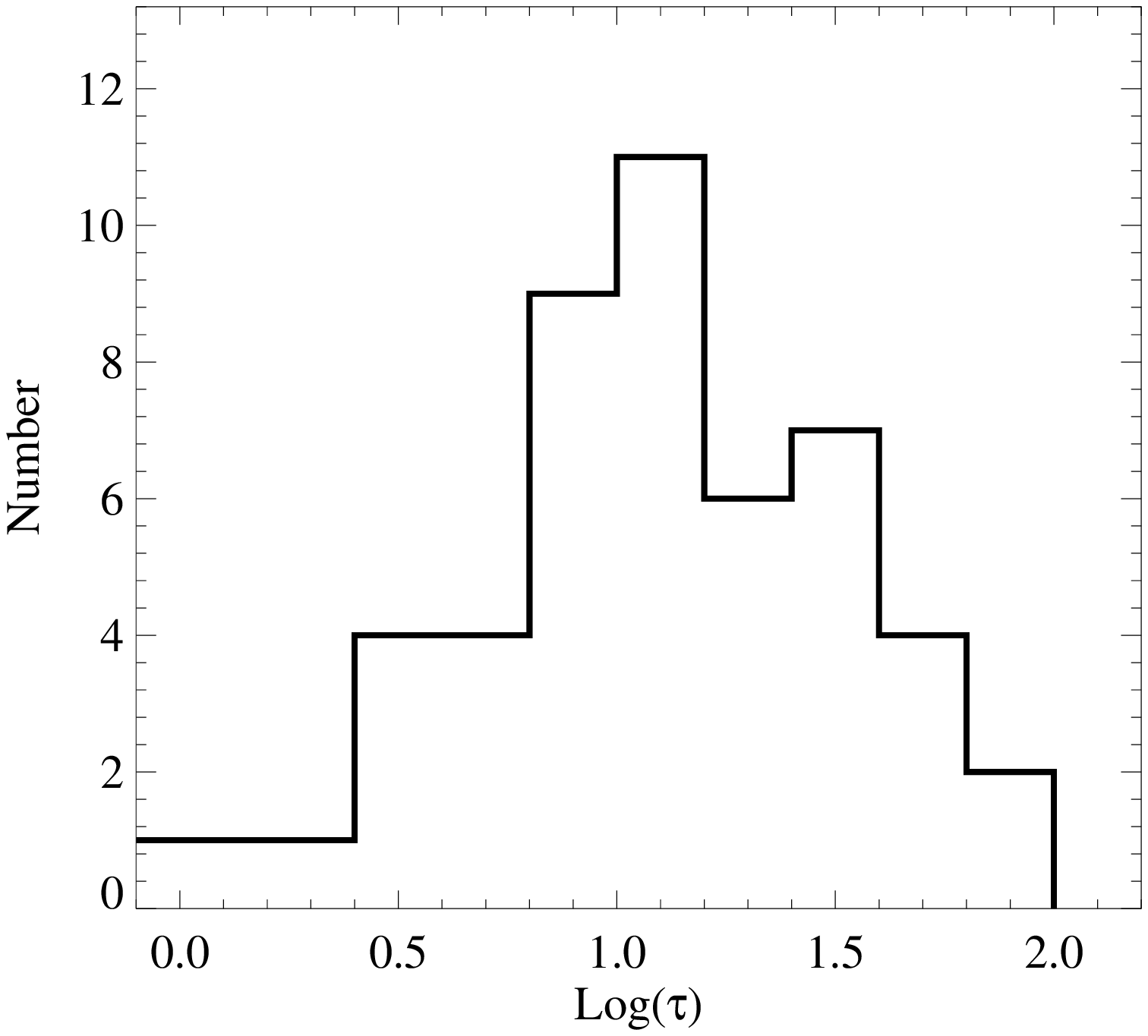}{0.45\textwidth}{(c)}
     \fig{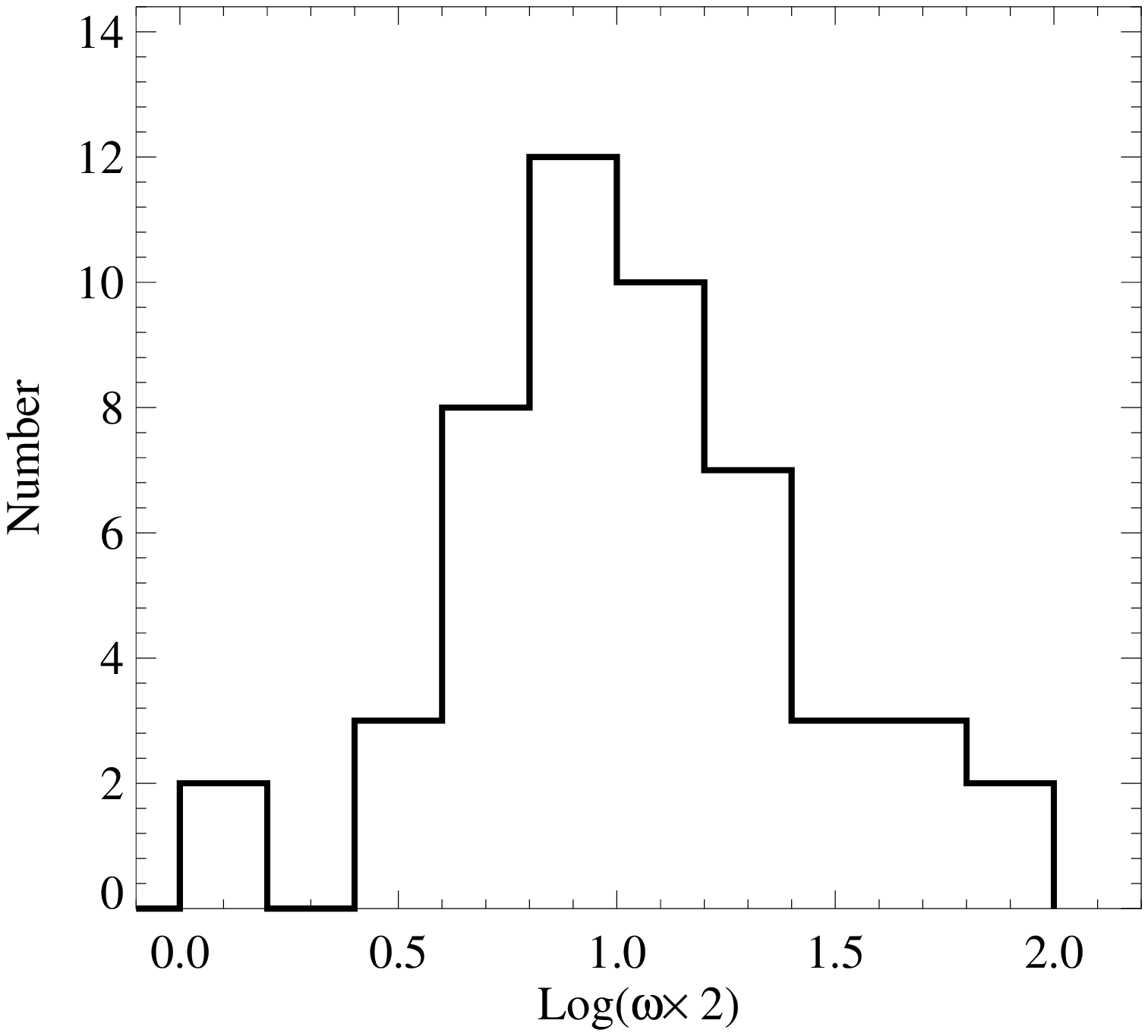}{0.45\textwidth}{(d)}
  			}   
\gridline{ \fig{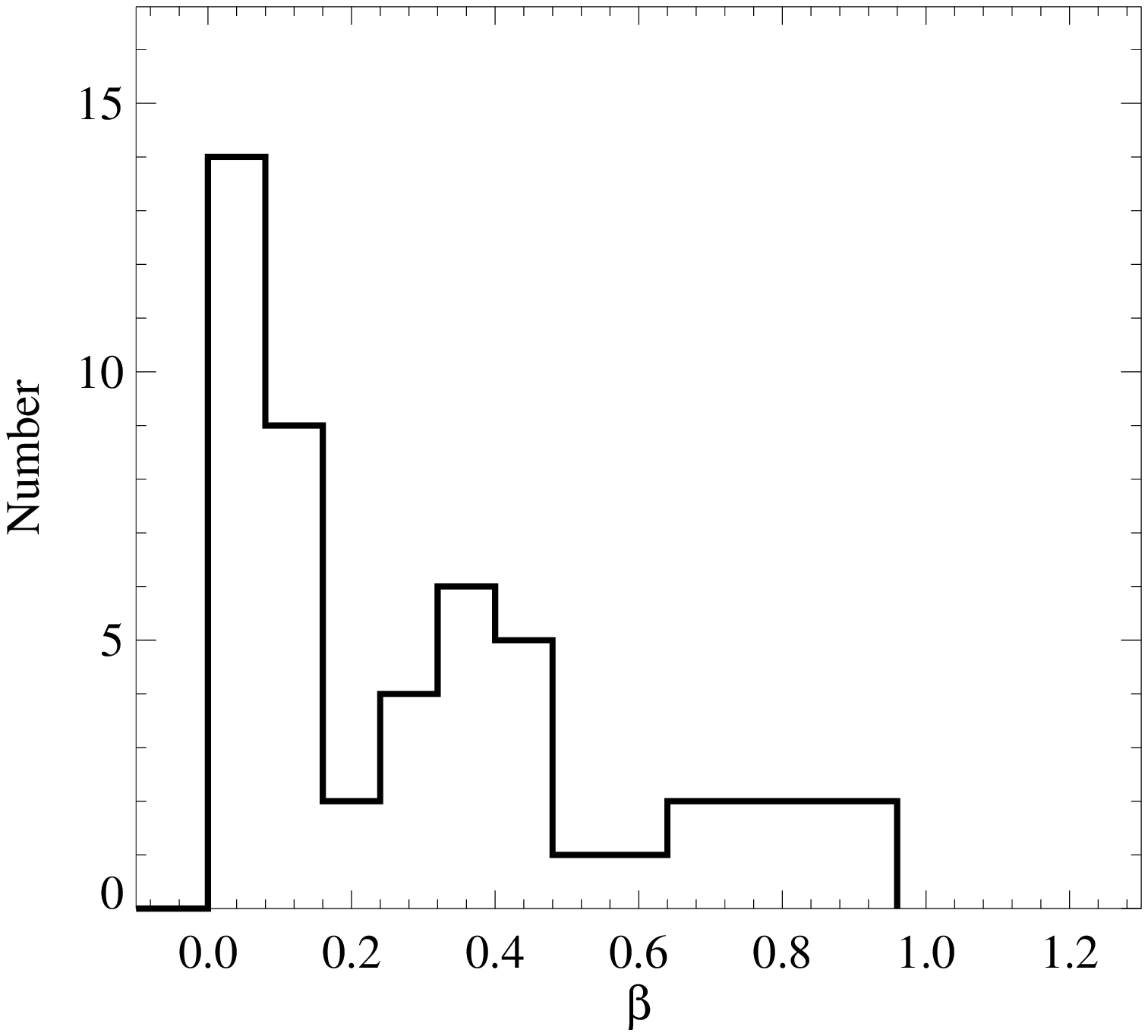}{0.45\textwidth}{(e)}
			\fig{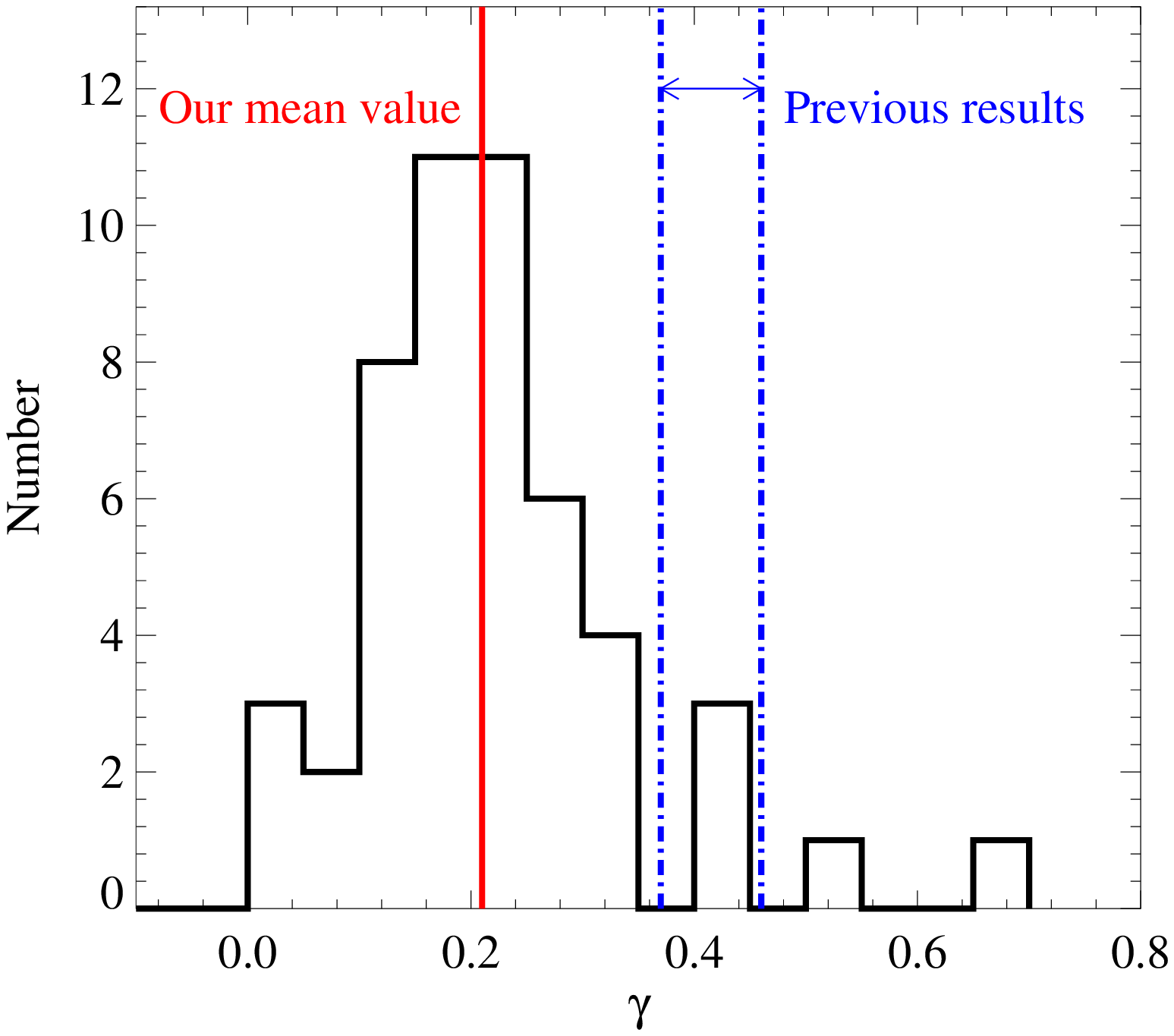}{0.45\textwidth}{(f)}
  			} 
\caption{Distributions of the burst duration $T_{90}$, the initial arrival time $t_0$, the ``limiting'' spectral lag $\tau$ , the ``limiting'' half width $\omega$, the power-law indices $\beta$ and $\gamma$ for the 50 GRBs in our sample. As shown by the panel~(f), the previously well-studied power-law index $\gamma$ was proposed to be between $\sim$ 0.37 and $\sim$ 0.46 (marked by two vertical blue dash-dotted lines) \citep[e.g.,][]{1995ApJ...448L.101F,1996ApJ...459..393N,2005ApJ...627..324N}, while our new results (in black histogram) have a mean value of $\sim$0.21 (marked by the vertical red solid line).  \label{fig:dist}}
\end{figure}

The effects of spectral lag has been manifested as both the delay of the peak arrival time and the broadening of the pulse width in a lower energy channel. These two phenomena appear to be closely connected. The best-fitting functions given by Equations~(\ref{eq:powerlaw1}) and (\ref{eq:powerlaw2}) are shown by the red curves in the right panels in Figure Set~\ref{fig:sample}.
To well constrain the values and uncertainties of each parameters in the power-laws, we have adopted the efficient Nested Sampling Monte Carlo algorithm in the framework of Bayesian analysis, used in the generic package PyMultiNest, as firstly applied to X-ray spectral analysis \citep{2014A&A...564A.125B}. The best-fitting parameters of $t_0$, $\tau$ , $\beta$, $\omega$ and $\gamma$ are provided in Table~\ref{tab:sample}. The distributions for these parameters together with the T$_{90}$ in our sample are shown in Figure~\ref{fig:dist}. 

The median value of $t_0$ is -1.03~s, which can be understood as the mean value of the limiting initial occurrence time of the GRB pulse. The median value of $T_{90}$, $\tau$ and $\omega$ is 12.4~s, 12.6~s and 5.2~s, respectively. There is a clear trend that the limiting spectral lag $\tau$ is correlated with the duration $T_{90}$ as shown in the Figure~\ref{fig:correlation}(a). Meanwhile, $\omega$ is also correlated with the duration $T_{90}$ as shown in Figure~\ref{fig:correlation}(c), and $\tau$ is correlated with the twice of $\omega$ as shown in Figure~\ref{fig:correlation}(e). These correlations suggest that $T_{90}$ is a modestly good indicator of the intrinsic duration of the GRB pulses in single pulse events. The real interesting correlation is between the pulse width and the spectral lag. The correlation between the spectral lag and the $T_{90}$ comes as a consequence. A similar correlation between the pulse width and the spectral lag (though, defined slightly differently) was also found in bright X-ray flares, suggesting a common origin of prompt emission and the X-ray flares \citep[e.g.,][]{2010MNRAS.406.2149M}. On the contrary, neither $\beta$ nor $\gamma$ appears to be correlated with the duration $T_{90}$ as shown in Figure~\ref{fig:correlation} (b) \& (d).

The connection between the delay of the peak arrival time and the broadening of the pulse width can also be revealed by the similar median values of $\beta=0.27$ and $\gamma=0.21$.  While $\gamma$ tends to have a unimodal distribution which tops at $\sim$ 0.2, $\beta$ tends to have a bimodal distribution which tops towards $\sim$ 0.1. This bimodal distribution of $\beta$ is also hinted by the heart-shaped correlation between $\tau$ and $\beta$ as shown in Figure~\ref{fig:correlation}(g), where two distinction components, i.e., an anti-correlation between $\tau$ and $\beta$ at $\beta< 0.3$ and a correlation at $\beta> 0.3$ are apparently present. These two components form the two peaks at $\beta \sim 0.1$ and $\beta \sim 0.4$ shown in Figure~\ref{fig:dist}(e). Figure~\ref{fig:correlation}(g) also indicates an intriguing trend that the cases with both $\tau \sim 0$ and $\beta \sim 0$ do not exist. For a comparison, an apparent correlation between $\omega$ and $\gamma$ is shown in Figure~\ref{fig:correlation}(h). The latter indicates that the width of narrower pulses are less dependent on the photon frequencies.
 
It is worth mentioning that, a mean value of the power-law index $\gamma=0.21$ appear to be inconsistent with the previously proposed values of $\gamma\sim 0.4$ for BATSE and Swift GRBs which were all based on analyses of average pulse width as determined either by the CCF or by specific fits of the light curves of individual pulses in given broad energy channels \citep{1995ApJ...448L.101F,1996ApJ...459..393N,2005ApJ...627..324N}. Recently, the energy dependence of minimum variability timescales has also been studied for Fermi/GRM bursts in four energy channels, and the power-law index has been found to be between 0.53 and 0.97 \citep{2015ApJ...811...93G}. A value of $\gamma\sim 0.4$ has also been found for X-ray flares \citep{2010MNRAS.406.2113C,2010MNRAS.406.2149M}. Alternatively, the recent work on the extremely bright GRB 130427A detected by Fermi/GBM evaluated a value of $\gamma=0.27\pm 0.03$ \citep{2014Sci...343...51P} which is consistent with our results here. The energy dependence of the pulse width with a power-law index of $\gamma \sim 0.5$ might indicate the signature of synchrotron cooling \citep[e.g.,][]{1998ApJ...493..708K,1998ApJ...508..752C}. On the contrary, a power-law index of $\gamma \sim 0.25$ might favor the dominance of the SSC processes \citep[e.g.,][]{1998ApJ...501L.157D}, the relativistic curvature effect \citep[e.g.,][]{2005MNRAS.362...59S}, or the Doppler effect \citep[e.g.,][]{2004ApJ...617..439Q}. However, as discussed in recent works, these models still have difficulties in fully accounting for the observational features, which instead indicate a combination of several specific constraints on the radiation mechanism \citep[e.g.,][]{2016ApJ...825...97U}. It was proposed that the spectral lag could be produced by the combination of spectral evolution and the curvature effect \citep{2011AN....332...92P}.

Even though short GRBs generally show negligible spectral lags \citep{2006ApJ...643..266N} or even ``negative'' lags \citep{2006MNRAS.367.1751Y}, the only short burst in our sample, i.e., 140209313 ($T_{90}=1.41$~s), does show an apparent spectral lag of $\tau=1.28$~s that is comparable to its duration as shown by the blue data points in Figure~\ref{fig:correlation}. The correlations between the spectral lag or the pulse width and the duration apply nicely for both short and long bursts in our sample, suggesting some similarity or connection between the two classes as suggested by previous works \citep[e.g.,][]{2009A&A...496..585G,2011ApJ...738...19S,2012ApJ...750...88Z}. While the values of $\beta$ and $\gamma$ for 140209313 are similar to those of the long bursts, it is, however, outlier in the $\omega$-$\gamma$ diagram as shown in Figure~\ref{fig:correlation}(h). Another special burst is 091010113 as shown by the purple data point. Although, 091010113 has a long duration of $T_{90}=5.95$~s as listed in the online catalogue, its major pulse (as shown in our Figure Set~\ref{fig:sample}) lasts obviously less than 2~s. There two special bursts are also located at the bottom of heart shape in the $\tau$-$\beta$ diagram in the panel~(g) of Figure~\ref{fig:correlation} or at the bottom right corner in the $\omega$-$\gamma$ diagram in the panel~(h) of Figure~\ref{fig:correlation}, suggesting a potential subgroup. 

\section{``Hard-to-Soft'' or ``Hardness-Intensity Tracking''?} \label{sec:hardness}

The peak energy E$_{\rm p}$ has been considered as an important spectral parameter of GRB pulses and found to be in correlation with many measured quantities, such as the isotropic total energy $E_{\rm \gamma,iso}$ \citep{2002A&A...390...81A}. However, none of the four newly-measured parameters related with the spectral lag shows any clear correlation with $E_{\rm p}$ as shown by Figure~\ref{fig:noncorrelation}, where the values of $E_{\rm p}$ have been adopted as the peak energy of a Band function fit to the single spectrum over the duration of the burst provided by the official GBM online burst catalogs \citep{2014ApJS..211...12G,2014ApJS..211...13V}. Given the significant spectral evolution in GRB pulses, time-resolved spectral analysis is a key to crack the underlying radiation mechanism. A recent hot topic is to distinguish the HTS and HIT patterns in the evolution of the hardness of the spectra. 

Instead of using the peak energy $E_{\rm p}$, in this work we directly use the hardness ratio as an indicator of the spectral properties of the GRBs. The hardness ratio is defined as the ratio between the two light curves in a harder and a softer energy channels, respectively, which are determined here for a preliminary result by evenly cutting the full energy channel in half in the logarithmic scale. This hardness ratio is easily available and is not subject to the uncertainties in determining $E_{\rm p}$ for a Band function fitting. The latter might has some statistical issues as suggested by the previous works \citep[e.g.][]{2016ApJ...821...12P}. Our results of the hardness ratio have been shown in the thick blue histogram in Figure~\ref{fig:sample}. For almost all the bursts in our sample, the hardness ratio shows a uniform pattern including both soft-to-hard and hard-to-soft phases, and has a very similar temporal profile as the total light curve. In a few of them, say, 130630272, the hardness ratio is nicely consistent with the total light curve clearly showing an HIT pattern. Most of the ones that cannot be considered as strictly HIT instead show a delay between the temporal profile of the hardness ratio and the total light curve. Strictly speaking, the majority of them seem to differ with either the HTS or HIT patterns. For example, GRBs 081224887, 090809978, 100612726, and 110817191 were categorized as HTS pattern based on analysis of the temporal evolution of the peak energy $E_{\rm p}$ \citep[e.g.,][]{2012ApJ...756..112L}. However, an initial ``soft-to-hard'' phase seem to be indispensable in these four cases as shown here. A very similar pattern between the hardness ratio and the total light curve can be perceived for almost all the bursts in our sample. Pure HTS pattern seems to be disfavored for most of the bursts in our sample. 

To further illustrate the discrepancy in the pattern of the spectral evolution between the previous analyses utilizing the peak energy $E_{\rm p}$ and our analysis utilizing the hardness ratio, we made a direct comparison of the results based on these two spectral quantities for the four previously-proposed HTS bursts mentioned above (i.e., 081224887, 090809978, 100612726, and 110817191). First, we reproduced the spectral analyses for the evolution of the peak energy $E_{\rm p}$ using the software package GBM RMFIT tool (version 4.3pr2) \footnote{http://fermi.gsfc.nasa.gov/ssc/data/analysis/rmfit/} and the results are shown in the top panels in Figure~\ref{fig:hardness} for each burst. We adopt the Bayesian blocks (BBs) as the binning method \citep{2013ApJ...764..167S} which is considered more accurate in reproducing the intrinsic spectral evolution pattern in highly variable GRB light curves as suggested by \cite{2014MNRAS.445.2589B}. The adopted time bins are listed in the second column of Table~\ref{tab:sample2}. We can reproduce the HTS pattern in the measure of $E_{\rm p}$, which is consistent with the previous result. For a straight comparison, we rebin the light curve in the same time bins and calculate the hardness ratio as introduced above. The results are shown in the bottom panels in Figure~\ref{fig:hardness} for each burst. Dramatically, the four supposedly HTS bursts determined by $E_{\rm p}$ appear to have an semi-HIT pattern, though not very statistical significant in GRBs 100612726 and 110817191. This suggests that an obvious ambiguity may exist in distinguishing HTS and HIT behaviors. The biggest difference is in the first time bin, which is understandable since the pulse is weak and the time bin is narrow therefore the uncertainty in determining the value of $E_{\rm p}$ is the largest.

More importantly, the evolution behavior of the spectral hardness for most GRBs in our sample is neither HTS nor HIT. An obvious time delay between the hardness evolution pattern and the light curve exists. To study the property of this hardness-intensity time delay $\Delta t_{\rm HR}$, we calculate the CCF between the evolution curve of hardness ratio (the thick blue curve in each left panel of Figure Set~\ref{fig:sample}) and the total light curve (the thick black curve in each left panel of Figure Set~\ref{fig:sample}) using the same method as described in \citet{2012ApJ...748..132Z}. The measured values of $\Delta t_{\rm HR}$ are listed in the last column of Table~\ref{tab:sample}. The origin of this time delay can be revealed by the correlation between $\Delta t_{\rm HR}$ and the spectral lag $\tau$ as shown in Figure~\ref{fig:HITlag}. The best fit yields
\begin{equation}
\Delta t_{\rm HR} =0.29 \times \tau ^{0.64 \pm 0.01} ,\label{eq:HITlagcor}
\end{equation}
with a Spearman's rank correlation coefficient $r=0.67$ and a chance probability $P=1.0\times 10^{-7}$. The suggests that the spectral evolution behavior (hardness vs. light curve) is closely related with the spectral lag phenomenon. Given an intrinsic time delay between the evolution of hardness and light curve, neither HTS or HIT is an appropriate categorization of the spectral evolution behavior for most GRBs. This time delay is an indication of the spectral lag. As a consequence, only the bursts with negligible spectral lag would exhibit a pure HIT behavior. The bursts with a large spectral lag would more likely exhibit an HTS behavior. Most GRBs have neither HIT nor HTS behaviors, yet form a major and continuous transition between HIT and HTS categories. Since the spectral lag is well correlated with the pulse width as found above, short pulses tend to show an HIT behavior and long pulses tend to show an HTS behaviors. As shown in Figure~\ref{fig:HITlag}, The only two short GRBs 091010113 and 140209313 that have negligible $\Delta t_{\rm HR}$ are well located in the lower left HIT corner, and the four previously proposed HTS GRBs with longer pulse widths all have a moderate $\Delta t_{\rm HR}$. 
 
\section{Discussions and conclusion} \label{sec:discussion}

In this work, we revisited the measurement of GRB spectral lag between a group of consecutive energy channels instead of two given energy channels. The idea of splitting the light curves into more consecutive energy channels was triggered by the studies of pulsars in the radio frequencies, even though the physics behind them are totally different. For pulsars there is a quadratic frequency dependence on the arrival time ($\Delta t_{\rm peak} \propto \nu^{-2}$) as is typical for the propagation effect in cold plasma along the line of sight, known as the dispersion \citep[e.g.,][]{2007Sci...318..777L}. Meanwhile, the significant pulse width broadening with a quadratic frequency dependence ($\propto \nu^{-4}$) is consistent with the Kolmogorov-like spectrum due to interstellar scattering \citep{1976ApJ...206..735L}. On the other hand, the spectral lag for GRBs in the MeV energies is still an open issue, which more likely has an internal origin due to the radiation mechanisms. Based on a systematical analysis on 50 single-pulsed GRBs,  we find that the arrival time of the GRB pulse is universally anti-correlated with the photon energy, showing an intrinsic soft spectral lag (i.e., softer photons come later) under $\sim$~800 keV as observed by the NaI detectors of Fermi/GBM. By investigating the pulse profiles in multiple consecutive energy channels, we can determine an intrinsic spectral lag $\tau$ and pulse width $\omega$ for each burst independent on energy channel selection. 

The spectral lag $\tau$ and pulse width $\omega$ are found well correlated with each other, which may favor the relativistic geometric effects previously proposed, e.g., the spectral lag is due to the fact that the observer is looking at the increasing latitudes with respect to the line of the sight with time \citep[see][for a discussion and the references therein]{2009ApJ...703.1696Z}. We can, for the first time, evaluate the energy-dependency for the pulse arrival time by the power-law index $\beta$, which is widely distributed between $\sim$~0.02 and 0.9 with a mean value of $\sim$~0.27. The pulse width is also found universally anti-correlated with the photon energy (i.e., softer pulses have a wider width) which is consistent with previous studies. The power-law index $\gamma$ for the energy-dependency of the pulse width is also widely distributed between $\sim$~0.02 and 0.7, but with a mean value of $\sim$~0.21 which is distinct from previous studies \citep[e.g.,][]{1995ApJ...448L.101F,1996ApJ...459..393N,2005ApJ...627..324N}. Our new result on the power-law index $\gamma$ would also favor the relativistic geometric effects. There should also be a caution that, for some cases, e.g., GRB 100324172, that the anti-correlations of the energy dependence could not be well constrained, potential systematic uncertainties may exist and probably translate into a very low value of $\gamma$ (or $\beta$ in some other case).

It is worth to mention that, our work reveals some interesting mutual correlations between the burst duration T$_{90}$, the spectral lag $\tau$, and the pulse width $\omega$. There might also exist a correlation between the power-law indices $\beta$ and $\gamma$ as shown in Figure~\ref{fig:correlation}(f), although with a mild Spearman's rank correlation coefficient $r$ of 0.4086 and the corresponding chance probability $P$ of 0.0032. Since the physical model that is fully consistent with the data is still missing, we may provide some hints about the connection between the energy-dependencies of both the arrival time and the pulse width. As a simple demonstration, we can show that, given the energy dependency of the arrival time fully described in Equations~(\ref{eq:powerlaw1}), the energy dependency of the pulse width described in Equations~(\ref{eq:powerlaw2}) might be a natural consequence. Let us assume that, the energy dependence of the peak arrival time (i.e. Equation~(\ref{eq:powerlaw1})) is a realization of the intrinsic radiation mechanism which applies to each emitted photon. Now, we consider a very simple pulse, which consists of only two photons with different energies $E_1$ and $E_2$ ($ E_1 < E_2 $), respectively. According to Equation~(\ref{eq:powerlaw1}), they each arrive at the time $t_1$ and $t_2$ by 
\begin{equation}
t_1 = t_0+\tau\times ({E_1\over 1\,{\rm keV}})^{-\beta},\label{eq:t1}
\end{equation}
and
\begin{equation}
t_2 = t_0+\tau\times ({E_2\over 1\,{\rm keV}})^{-\beta},\label{eq:t2}
\end{equation} 
Now we collect these two photons in the energy channel [$E_1$, $E_2$], and measure the width $W$ of this simple pulse formed by the only two photon, which is the time difference between $t_1$ and $t_2$ by
\begin{eqnarray}
W = t_1 -t_2 & = & \tau\times\left[ ({E_1\over 1\,{\rm keV}})^{-\beta} - ({E_2\over 1\,{\rm keV}})^{-\beta} \right],\\
			& \approx & \tau\times ({E_1\over 1\,{\rm keV}})^{-\beta}. \label{eq:t1t2}
\end{eqnarray}
Comparing Equations~(\ref{eq:powerlaw2}) and (\ref{eq:t1t2}) suggests that $\tau \approx 2 \times \omega$ and $\beta \approx \gamma$, which are consistent with the results as shown in Figure~\ref{fig:correlation}(e)\& (f). This means that the energy dependence of the characteristic arrival time of a group of photons should be inevitably reflected in a similar energy dependence of the pulse width of the same group of photons. We doubt that one can be able to distribute a group of photons that have significantly energy-dependent arrival times yet can also have an energy-independent pulse width. Based on our analysis above, therefore, we propose that the energy dependency of the arrival time of the emitted photons (Equation~(\ref{eq:powerlaw1})) is more likely the internal and intrinsic radiation mechanism and the energy dependency of the pulse width (Equation~(\ref{eq:powerlaw2})) is its external and natural/inevitable consequence. Anyway, a complete theoretical framework that can fully account for the energy dependencies of both arrival time and pulse width is still missing. 

\begin{figure}
\figurenum{2}
\epsscale{0.9}
\plottwo{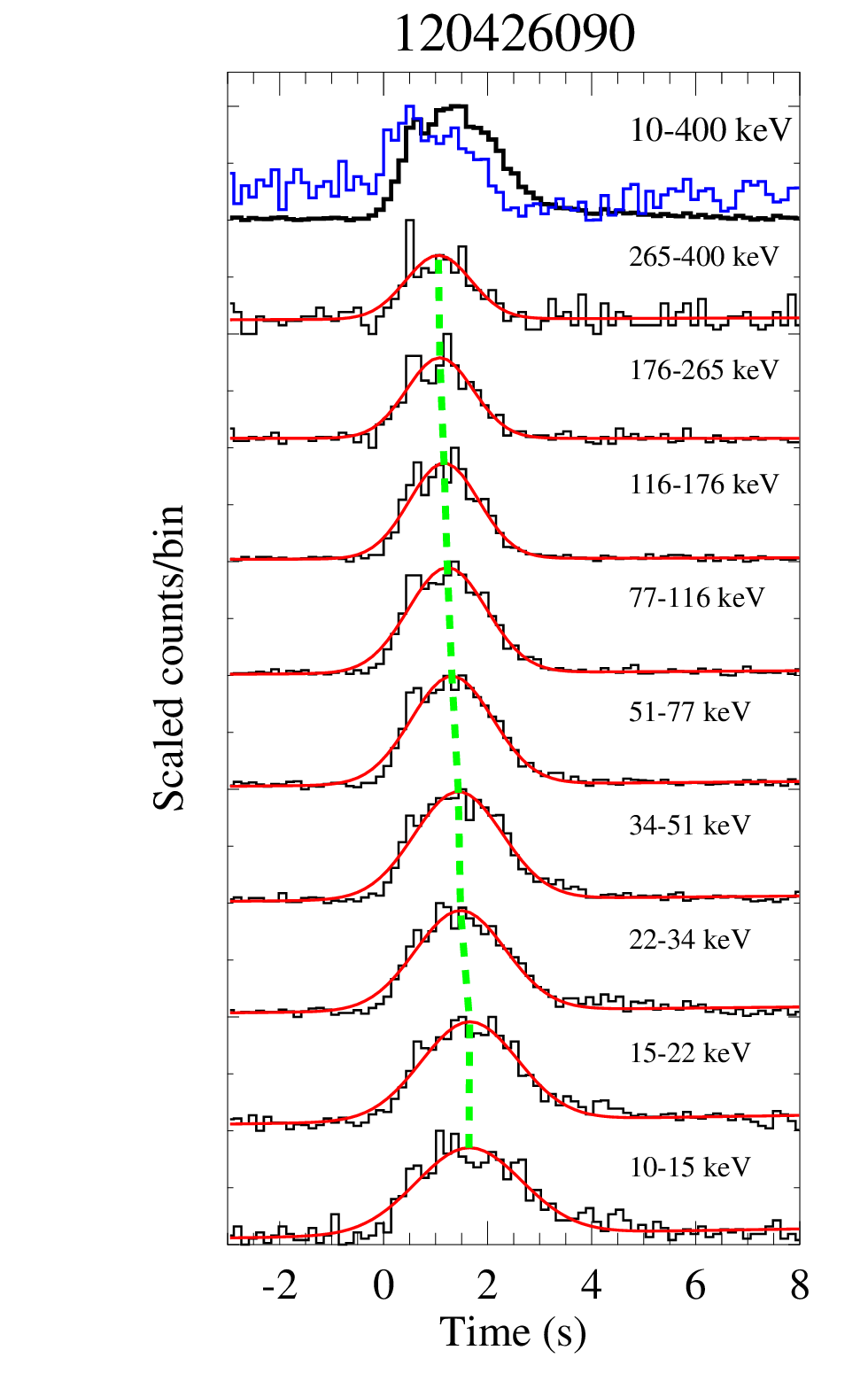}{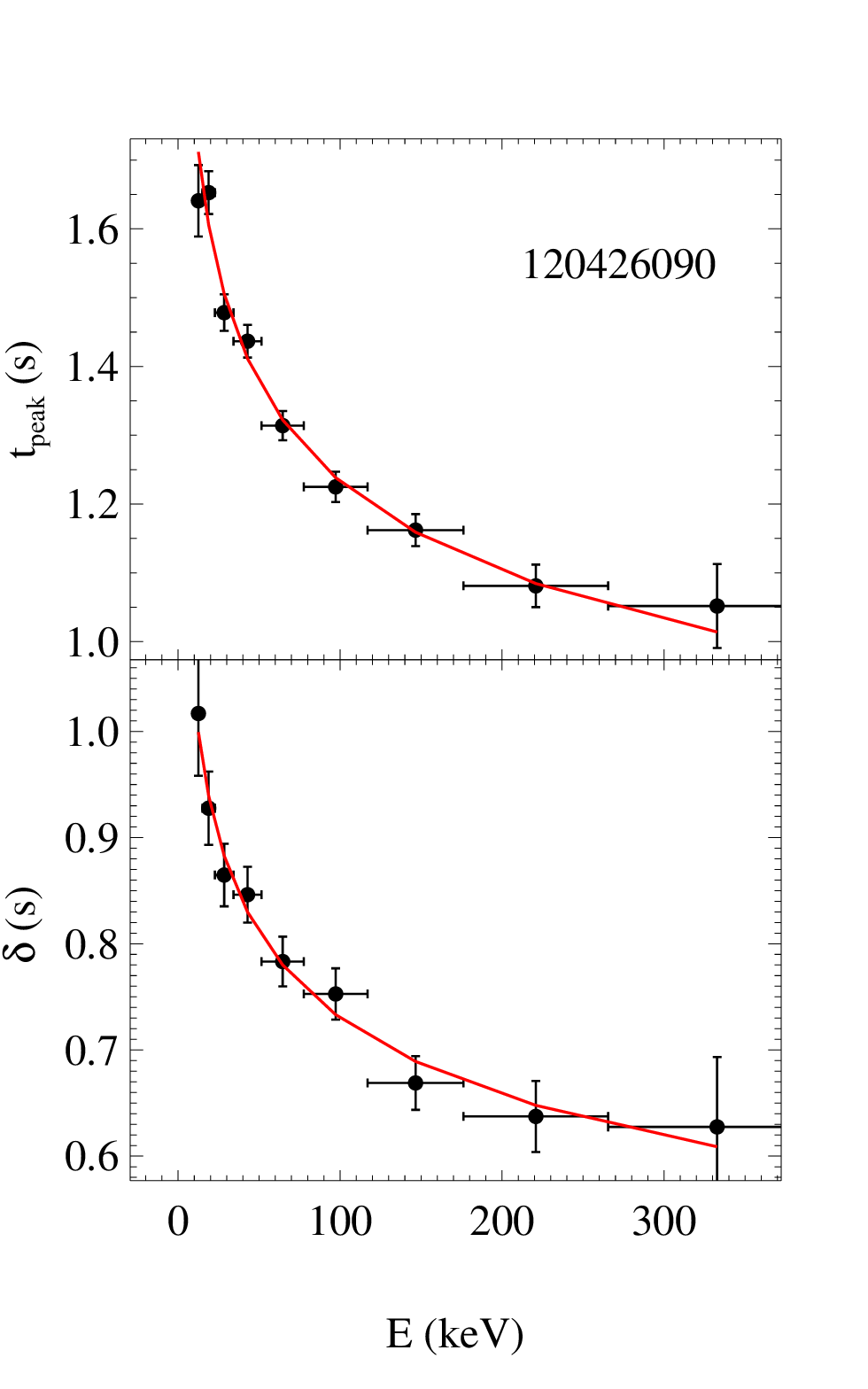}
\caption{GRB~120426090 as an example in our sample. Left panel shows the light curve of GRB~120426090 in nine consecutive energy channels in thin black histogram, over-plotted with the best-fit Gaussian functions in red curves. Additional first-order polynomial functions have been adopted to subtract the backgroud. The peak points of each Gaussian function are joined by a green dashed curve for a visual aid to manually check the effect of spectral lag. On top of the nine light curves, the normalized hardness ratio in thick blue histogram is plotted together with the total light curve in thick black histogram. Right panel shows the energy dependencies of the peak arrival time $t_{\rm peak}$ (right top) and the RMS width $\delta$ (right bottom) of the pulses, respectively. The best-fit functions given by Equations~(\ref{eq:powerlaw1}) and (\ref{eq:powerlaw2}) are shown by the red curves for the right top and right bottom panels, respectively. [The complete Figure Set 2 (100 images for 50 bursts) is available in the online journal.]}\label{fig:sample}
\end{figure}

We used a symmetric Guassian profile to describe the single-pulse profiles of the bursts in our sample. Given the potential asymmetry of the pulses, there might be possible systematical uncertainties of the pulse peak times. Since we had some difficulties in fitting all the lightcurves in our sample with any exact function provided in the literature, we further adopted CCF method to check the delay in the peak time between two consecutive energy channels and compared it with our results based on Guassian fitting. Our CCF method has been applied and explained in \cite{2012ApJ...748..132Z}. The comparison for each burst is shown in Figure~\ref{fig:compare}. The same time ranges for each burst as shown in in the left panels of Figure Set~\ref{fig:sample} were used to calculate the CCF lags. All the uncertainties of CCF lags in this work were estimated by Monte Carlo simulation as described in \citet{2012ApJ...748..132Z} and the uncertainties of Gaussian peak-time lags were estimated by error propagation of the peak times. Most of the results from different methods are consistent with each other. Though a few cases, say, GRB 100707032 and 150314205, do reveal a systematical overestimation because of the pulse confusion as can be perceived by the fittings for the light curves in the lower energy channels shown in Figure Set~\ref{fig:sample}. Given that their numbers are few, the general results for the sample in this work have not been significantly affected by our current choice of the fitting function. Anyway, the intrinsic shape of GRB pulses is still an open issue. A more complicated fitting scheme \citep[e.g.,][]{2005ApJ...627..324N} for the pulse profile should be explored for the work in the future.

As an important result in this work, we found a new clue to solve the controversy over the spectral evolution of GRB pulses, i.e., the so-called Hard-To-Soft (HTS) and Hardness-Intensity Tracking (HIT) behaviors. We use the hardness ratio instead of the peak energy $E_{\rm p}$ as the characteristic spectral parameter since the later is subject to uncertainties in the very early stage of the pulse. We find that most of the GRB pulses exhibit neither HTS nor HIT behaviors. Instead, they show a behavior close to HIT but with a noticeable time delay. This time delay $\Delta t_{\rm HR}$ is well correlated with the spectral lag $\tau$ that we have measured in this work. A GRB pulse with negligible spectral lag would appear to have an HIT behavior. On the other hand, a GRB pulse with significant spectral lag would appear to have an HTS behavior, which, in fact, is not a genuine behavior given that there is always a very short soft-to-hard phase at the very beginning. Even though most GRBs are highly variable with multiple pulses, single pulses seem to be fundamental elements of GRB prompt emission \citep[e.g.,][]{2008ApJ...677L..81H}. Our results on single-pulse GRBs can provide important insights into the nature of GRB pulses.

\acknowledgments

This work was supported in part by the the National Basic Research Program (``973" Program) of China (Grant No. 2014CB845800), the National Natural Science Foundation of China (Grants Nos.  11103083, 11433009, 11673068, and 11533003) and the Strategic Priority Research Program “Multi-wavelength Gravitational Wave Universe” of the Chinese Academy of Sciences (No. XDB23000000). LS acknowledges the supported by the Joint NSFC-ISF Research Program (No. 11361140349), jointly funded by the National Natural Science Foundation of China and the Israel Science Foundation. BBZ acknowledges the support by the Spanish Ministry Projects AYA 2012-39727-C03-01 and AYA2015-71718-R. XFW acknowledges the support by the Youth Innovation Promotion Association (2011231) and the Key Research Program of Frontier Sciences (QYZDB-SSW-SYS005) of the Chinese Academy of Sciences. DX acknowledges the support by the One-Hundred-Talent Program of the Chinese Academy of Sciences (CAS). Part of this work used BBZ's personal IDL code library ZBBIDL and personal Python library ZBBPY. The computation resources used in this work are owned by Scientist Support LLC.  This research has made use of NASA's Astrophysics Data System Bibliographic Services. 

\bibliographystyle{aasjournal}

\begin{deluxetable}{lrcrrrrrcc}
\tablecaption{Our sample of 50 GRBs detected by GBM and results on the spectral lag.\label{tab:sample}}
\tablewidth{0pt}
\tablehead{
\colhead{Burst} & \colhead{$T_{90}$~(s)\tablenotemark{a,b}} & \colhead{Bin~(s)} & \colhead{$t_0$~(s)} & \colhead{$\tau$~(s)} & \colhead{$\beta$} & \colhead{$\omega$~(s)} & \colhead{$\gamma$} & \colhead{NaI\tablenotemark{c}} & \colhead{$\Delta t_{\rm HR}$~(s)}
}
\startdata
081125496 & 9.28$\pm$0.61 & 0.46 & -15.770$_{-0.142}^{+4.420}$ & 23.560$_{-4.066}^{+0.225}$ & 0.047$_{-0.003}^{+0.035}$ & 8.456$_{-0.746}^{+0.932}$ & 0.336$_{-0.022}^{+0.026}$ & na & 2.521$\pm$0.126 \\
081224887 & 16.45$\pm$1.16 & 0.82 & -11.103$_{-0.116}^{+3.040}$ & 18.545$_{-2.713}^{+0.131}$ & 0.058$_{-0.003}^{+0.033}$ & 4.135$_{-0.408}^{+0.559}$ & 0.171$_{-0.022}^{+0.028}$ & n6 & 1.174$\pm$0.125 \\
090719063 & 11.39$\pm$0.47 & 0.57 & 2.926$_{-0.632}^{+0.609}$ & 18.732$_{-2.311}^{+5.054}$ & 0.488$_{-0.085}^{+0.127}$ & 8.557$_{-0.621}^{+0.755}$ & 0.246$_{-0.019}^{+0.022}$ & n8 & 1.678$\pm$0.131 \\
090804940 & 5.57$\pm$0.36 & 0.28 & -16.344$_{-0.258}^{+5.194}$ & 20.837$_{-4.702}^{+0.340}$ & 0.022$_{-0.003}^{+0.090}$ & 4.735$_{-0.393}^{+0.489}$ & 0.238$_{-0.021}^{+0.025}$ & n5 & 0.828$\pm$0.086 \\
090809978 & 11.01$\pm$0.32 & 0.55 & -4.747$_{-6.670}^{+4.876}$ & 14.166$_{-3.057}^{+6.186}$ & 0.110$_{-0.047}^{+0.130}$ & 6.128$_{-0.527}^{+0.656}$ & 0.241$_{-0.021}^{+0.025}$ & n4 & 1.908$\pm$0.128 \\
090820027 & 12.42$\pm$0.18 & 0.62 & 34.182$_{-0.014}^{+0.177}$ & 12.640$_{-0.016}^{+58.098}$ & 0.786$_{-0.291}^{+0.291}$ & 4.985$_{-0.387}^{+0.507}$ & 0.142$_{-0.016}^{+0.020}$ & n2 & 1.157$\pm$0.151 \\
090922539 & 87.04$\pm$0.81 & 0.60 & 2.522$_{-0.031}^{+0.311}$ & 15.469$_{-1.196}^{+58.945}$ & 0.882$_{-0.013}^{+0.672}$ & 5.710$_{-0.941}^{+1.592}$ & 0.228$_{-0.044}^{+0.062}$ & n9 & 0.941$\pm$0.159 \\
091010113 & 5.95$\pm$0.14 & 0.05 & 2.219$_{-0.060}^{+0.060}$ & 1.224$_{-0.300}^{+0.559}$ & 0.374$_{-0.088}^{+0.158}$ & 0.760$_{-0.087}^{+0.124}$ & 0.339$_{-0.029}^{+0.037}$ & n3 & 0.102$\pm$0.009 \\
100324172 & 17.92$\pm$2.06 & 0.90 & -19.702$_{-0.295}^{+6.729}$ & 27.893$_{-6.158}^{+1.088}$ & 0.037$_{-0.008}^{+0.050}$ & 2.596$_{-0.295}^{+0.472}$ & 0.032$_{-0.027}^{+0.039}$ & n2 & 1.416$\pm$0.150 \\
100515467 & 10.62$\pm$1.43 & 0.53 & 0.922$_{-0.054}^{+0.054}$ & 1.764$_{-1.151}^{+32.415}$ & 0.359$_{-0.283}^{+0.477}$ & 2.085$_{-0.268}^{+0.415}$ & 0.212$_{-0.033}^{+0.045}$ & n7 & 0.256$\pm$0.068 \\
100528075 & 22.46$\pm$0.75 & 1.12 & 7.425$_{-0.139}^{+0.728}$ & 7.593$_{-0.118}^{+44.195}$ & 0.477$_{-0.052}^{+0.831}$ & 8.132$_{-0.662}^{+0.881}$ & 0.135$_{-0.020}^{+0.024}$ & n7 & 1.374$\pm$0.215 \\
100612726 & 8.58$\pm$3.21 & 0.43 & -19.597$_{-0.402}^{+6.497}$ & 26.430$_{-5.708}^{+0.683}$ & 0.029$_{-0.004}^{+0.096}$ & 2.607$_{-0.203}^{+0.274}$ & 0.088$_{-0.021}^{+0.026}$ & n4 & 2.656$\pm$0.117 \\
100707032 & 81.79$\pm$1.22 & 1.00 & 0.726$_{-0.190}^{+0.032}$ & 97.539$_{-8.153}^{+2.460}$ & 0.804$_{-0.060}^{+0.000}$ & 49.183$_{-5.219}^{+6.636}$ & 0.665$_{-0.022}^{+0.025}$ & n8 & 1.305$\pm$0.165 \\
101126198 & 43.84$\pm$1.75 & 0.50 & 9.244$_{-0.416}^{+0.416}$ & 4.837$_{-1.232}^{+64.333}$ & 0.085$_{-0.068}^{+2.734}$ & 6.185$_{-0.476}^{+0.615}$ & 0.104$_{-0.020}^{+0.024}$ & n7 & 1.749$\pm$0.159 \\
110301214 & 5.69$\pm$0.36 & 0.50 & 2.031$_{-2.158}^{+0.520}$ & 3.559$_{-1.317}^{+2.614}$ & 0.350$_{-0.242}^{+0.394}$ & 2.371$_{-0.173}^{+0.221}$ & 0.088$_{-0.020}^{+0.024}$ & n8 & 2.029$\pm$0.107 \\
110605183 & 82.69$\pm$3.08 & 1.00 & -7.452$_{-0.140}^{+9.839}$ & 25.367$_{-1.900}^{+19.457}$ & 0.149$_{-0.003}^{+0.452}$ & 16.274$_{-2.665}^{+5.620}$ & 0.310$_{-0.039}^{+0.066}$ & n2 & 2.459$\pm$0.237 \\
110721200 & 21.82$\pm$0.57 & 0.50 & -3.272$_{-0.143}^{+1.258}$ & 8.470$_{-0.848}^{+0.848}$ & 0.098$_{-0.009}^{+0.065}$ & 2.497$_{-0.242}^{+0.329}$ & 0.109$_{-0.024}^{+0.031}$ & n7 & 0.964$\pm$0.088 \\
110817191 & 5.95$\pm$0.57 & 0.30 & -3.319$_{-0.528}^{+1.153}$ & 8.290$_{-0.623}^{+0.623}$ & 0.114$_{-0.017}^{+0.070}$ & 3.971$_{-0.423}^{+0.591}$ & 0.299$_{-0.025}^{+0.032}$ & n9 & 0.717$\pm$0.059 \\
111009282 & 20.74$\pm$4.22 & 1.04 & -17.923$_{-1.835}^{+6.918}$ & 27.799$_{-6.463}^{+6.463}$ & 0.024$_{-0.007}^{+0.215}$ & 5.177$_{-0.496}^{+0.686}$ & 0.105$_{-0.027}^{+0.034}$ & n1 & 3.317$\pm$0.207 \\
111017657 & 11.07$\pm$0.41 & 0.55 & -0.226$_{-1.110}^{+1.447}$ & 7.894$_{-1.361}^{+1.361}$ & 0.078$_{-0.024}^{+0.171}$ & 4.466$_{-0.474}^{+0.676}$ & 0.184$_{-0.024}^{+0.031}$ & n6 & 0.991$\pm$0.101 \\
120119170 & 55.30$\pm$6.23 & 2.77 & 1.771$_{-3.838}^{+3.838}$ & 22.122$_{-5.626}^{+38.296}$ & 0.100$_{-0.078}^{+0.438}$ & 16.617$_{-2.092}^{+3.023}$ & 0.193$_{-0.034}^{+0.045}$ & nb & 3.775$\pm$0.439 \\
120426090 & 2.88$\pm$0.18 & 0.14 & -0.155$_{-3.346}^{+0.656}$ & 2.679$_{-0.303}^{+3.113}$ & 0.143$_{-0.099}^{+0.111}$ & 1.464$_{-0.086}^{+0.105}$ & 0.151$_{-0.015}^{+0.017}$ & n2 & 0.684$\pm$0.039 \\
120427054 & 5.63$\pm$0.57 & 0.50 & -1.847$_{-0.088}^{+0.845}$ & 8.166$_{-1.091}^{+0.229}$ & 0.153$_{-0.014}^{+0.080}$ & 2.731$_{-0.362}^{+0.564}$ & 0.162$_{-0.032}^{+0.043}$ & na & 0.953$\pm$0.079 \\
120625119 & 7.42$\pm$0.57 & 0.37 & 2.649$_{-0.145}^{+0.314}$ & 4.745$_{-0.743}^{+14.566}$ & 0.456$_{-0.093}^{+0.514}$ & 7.166$_{-0.788}^{+1.243}$ & 0.427$_{-0.026}^{+0.038}$ & n5 & 0.665$\pm$0.086 \\
120727681 & 10.50$\pm$1.64 & 0.53 & 1.764$_{-0.094}^{+0.094}$ & 2.965$_{-2.960}^{+36.079}$ & 0.200$_{-0.146}^{+3.338}$ & 6.622$_{-0.815}^{+1.352}$ & 0.247$_{-0.032}^{+0.046}$ & n2 & 1.198$\pm$0.203 \\
120919309 & 21.25$\pm$1.81 & 0.60 & 2.818$_{-0.125}^{+0.228}$ & 11.280$_{-2.432}^{+16.023}$ & 0.708$_{-0.104}^{+0.360}$ & 6.275$_{-0.605}^{+0.836}$ & 0.293$_{-0.023}^{+0.029}$ & n1 & 0.830$\pm$0.095 \\
121122885 & 7.94$\pm$0.57 & 0.40 & -19.883$_{-0.116}^{+5.070}$ & 34.420$_{-4.068}^{+1.431}$ & 0.081$_{-0.009}^{+0.038}$ & 4.030$_{-0.434}^{+0.655}$ & 0.106$_{-0.026}^{+0.035}$ & na & 1.607$\pm$0.101 \\
121223300 & 11.01$\pm$0.72 & 0.55 & 2.334$_{-0.641}^{+0.487}$ & 45.081$_{-17.034}^{+39.613}$ & 0.755$_{-0.182}^{+0.224}$ & 7.409$_{-1.193}^{+2.277}$ & 0.237$_{-0.040}^{+0.063}$ & n7 & 2.266$\pm$0.127 \\
130206482 & 11.26$\pm$1.95 & 0.50 & 2.064$_{-0.217}^{+0.217}$ & 8.709$_{-5.110}^{+30.579}$ & 0.617$_{-0.393}^{+0.393}$ & 3.412$_{-0.454}^{+0.788}$ & 0.199$_{-0.035}^{+0.053}$ & n1 & 1.345$\pm$0.123 \\
130325203 & 6.91$\pm$0.72 & 0.35 & 1.042$_{-0.443}^{+0.914}$ & 5.117$_{-1.736}^{+5.929}$ & 0.303$_{-0.072}^{+0.556}$ & 3.418$_{-0.442}^{+0.706}$ & 0.248$_{-0.031}^{+0.043}$ & n7 & 0.738$\pm$0.065 \\
130509078 & 24.32$\pm$3.59 & 1.22 & 2.433$_{-0.033}^{+0.422}$ & 14.821$_{-14.715}^{+33.794}$ & 0.827$_{-0.361}^{+0.361}$ & 7.527$_{-1.189}^{+2.358}$ & 0.259$_{-0.043}^{+0.070}$ & n9 & 0.604$\pm$0.282 \\
130518580 & 48.58$\pm$0.92 & 1.00 & 18.033$_{-0.538}^{+2.341}$ & 10.493$_{-2.069}^{+2.069}$ & 0.036$_{-0.014}^{+0.180}$ & 3.276$_{-0.253}^{+0.340}$ & 0.041$_{-0.018}^{+0.023}$ & n3 & 0.943$\pm$0.164 \\
130630272 & 17.15$\pm$0.57 & 0.86 & 5.245$_{-1.479}^{+0.345}$ & 73.604$_{-46.599}^{+12.555}$ & 0.893$_{-0.386}^{+0.071}$ & 17.024$_{-2.551}^{+5.267}$ & 0.264$_{-0.038}^{+0.065}$ & n4 & 0.886$\pm$0.209 \\
130701060 & 20.22$\pm$1.73 & 0.50 & 2.861$_{-0.069}^{+0.069}$ & 2.853$_{-2.820}^{+34.833}$ & 0.295$_{-0.233}^{+3.402}$ & 3.994$_{-0.724}^{+1.549}$ & 0.153$_{-0.045}^{+0.073}$ & na & 0.275$\pm$0.104 \\
130704560 & 6.40$\pm$0.57 & 0.50 & 1.229$_{-0.384}^{+0.349}$ & 10.478$_{-1.005}^{+2.006}$ & 0.448$_{-0.078}^{+0.103}$ & 4.876$_{-0.301}^{+0.358}$ & 0.247$_{-0.017}^{+0.019}$ & n4 & 5.186$\pm$4.424 \\
131014215 & 3.20$\pm$0.09 & 0.16 & -6.485$_{-0.126}^{+2.125}$ & 10.698$_{-2.065}^{+0.155}$ & 0.040$_{-0.006}^{+0.025}$ & 2.186$_{-0.130}^{+0.155}$ & 0.153$_{-0.017}^{+0.020}$ & na & 0.572$\pm$0.039 \\
131028076 & 17.15$\pm$0.57 & 0.86 & 5.962$_{-1.970}^{+0.942}$ & 13.255$_{-0.623}^{+2.759}$ & 0.305$_{-0.111}^{+0.111}$ & 9.094$_{-0.633}^{+0.775}$ & 0.211$_{-0.016}^{+0.018}$ & n2 & 1.625$\pm$0.151 \\
131216081 & 19.26$\pm$3.60 & 0.50 & -14.778$_{-1.151}^{+1.151}$ & 17.919$_{-12.889}^{+20.736}$ & 0.023$_{-0.010}^{+0.519}$ & 2.498$_{-0.497}^{+1.094}$ & 0.179$_{-0.047}^{+0.079}$ & n9 & 0.251$\pm$0.072 \\
131231198 & 31.23$\pm$0.57 & 1.56 & 20.476$_{-0.695}^{+0.815}$ & 35.391$_{-2.863}^{+5.885}$ & 0.460$_{-0.051}^{+0.084}$ & 20.614$_{-1.129}^{+1.374}$ & 0.331$_{-0.015}^{+0.017}$ & n3 & 8.519$\pm$0.534 \\
140209313 & 1.41$\pm$0.26 & 0.05 & 1.477$_{-0.082}^{+0.068}$ & 0.922$_{-0.067}^{+0.300}$ & 0.353$_{-0.099}^{+0.165}$ & 0.571$_{-0.056}^{+0.079}$ & 0.269$_{-0.023}^{+0.029}$ & na & 0.122$\pm$0.009 \\
140821997 & 32.51$\pm$1.64 & 1.00 & -19.073$_{-0.075}^{+13.702}$ & 56.024$_{-12.322}^{+0.034}$ & 0.032$_{-0.004}^{+0.059}$ & 11.621$_{-1.303}^{+1.823}$ & 0.201$_{-0.027}^{+0.034}$ & n5 & 3.551$\pm$0.296 \\
141028455 & 31.49$\pm$2.43 & 1.00 & 6.716$_{-8.459}^{+3.587}$ & 14.993$_{-0.696}^{+7.844}$ & 0.185$_{-0.103}^{+0.194}$ & 8.961$_{-0.806}^{+1.045}$ & 0.184$_{-0.022}^{+0.026}$ & n6 & 2.393$\pm$0.197 \\
150306993 & 18.94$\pm$1.15 & 0.95 & -0.424$_{-2.952}^{+2.193}$ & 27.668$_{-2.744}^{+20.100}$ & 0.358$_{-0.126}^{+0.238}$ & 23.230$_{-3.543}^{+5.790}$ & 0.413$_{-0.036}^{+0.049}$ & n4 & 1.716$\pm$0.166 \\
150314205 & 10.69$\pm$0.14 & 0.53 & -0.004$_{-2.038}^{+1.155}$ & 9.060$_{-0.149}^{+3.836}$ & 0.269$_{-0.115}^{+0.199}$ & 6.024$_{-0.611}^{+0.901}$ & 0.264$_{-0.024}^{+0.032}$ & n9 & 1.065$\pm$0.083 \\
150721242 & 18.43$\pm$0.57 & 0.92 & -4.904$_{-2.286}^{+1.778}$ & 51.275$_{-2.343}^{+3.957}$ & 0.368$_{-0.057}^{+0.060}$ & 35.815$_{-2.679}^{+3.347}$ & 0.544$_{-0.020}^{+0.023}$ & n7 & 11.371$\pm$1.728 \\
151021791 & 7.23$\pm$0.60 & 0.36 & 0.943$_{-0.033}^{+0.161}$ & 7.772$_{-0.849}^{+28.501}$ & 0.713$_{-0.042}^{+0.551}$ & 1.397$_{-0.175}^{+0.255}$ & 0.111$_{-0.029}^{+0.036}$ & na & 0.422$\pm$0.051 \\
151107851 & 139.01$\pm$6.45 & 1.00 & 8.162$_{-0.048}^{+0.885}$ & 11.509$_{-0.722}^{+47.452}$ & 0.474$_{-0.016}^{+0.650}$ & 23.531$_{-2.965}^{+4.388}$ & 0.427$_{-0.030}^{+0.039}$ & n9 & 1.568$\pm$0.239 \\
160101030 & 4.67$\pm$0.60 & 0.40 & -1.767$_{-0.096}^{+1.029}$ & 4.882$_{-0.828}^{+0.828}$ & 0.067$_{-0.018}^{+0.182}$ & 1.423$_{-0.132}^{+0.179}$ & 0.019$_{-0.027}^{+0.033}$ & n2 & 1.059$\pm$0.080 \\
160113398 & 24.58$\pm$0.26 & 1.23 & -14.304$_{-1.444}^{+13.188}$ & 54.794$_{-12.636}^{+1.614}$ & 0.030$_{-0.002}^{+0.041}$ & 8.164$_{-0.585}^{+0.703}$ & 0.180$_{-0.019}^{+0.021}$ & nb & 5.217$\pm$0.377 \\
160530667 & 9.02$\pm$0.18 & 0.45 & 1.513$_{-0.158}^{+1.207}$ & 6.642$_{-0.950}^{+0.223}$ & 0.070$_{-0.007}^{+0.060}$ & 3.393$_{-0.197}^{+0.226}$ & 0.112$_{-0.014}^{+0.015}$ & n2 & 0.913$\pm$0.084 \\
\enddata
\tablenotetext{a}{Data provided by the official GBM online burst catalogs \citep{2014ApJS..211...12G,2014ApJS..211...13V}.}
\tablenotetext{b}{All of the errors in this work indicate a confidence interval of $1\,\sigma$ uncertainty.}
\tablenotetext{c}{Only the data from the brightest NaI detector are used for analysis in this work.}
\end{deluxetable}

\clearpage
\begin{deluxetable}{lccc}
\tablecaption{Four supposedly HTS bursts.\label{tab:sample2}}
\tablewidth{0pt}
\tablehead{
\colhead{Burst Name} & \colhead{Time Bin~(s)} & \colhead{$E_{\rm p}$~(keV)} & \colhead{Hardness Ratio} \\
}
\startdata
081224887 & 0.001 $-$ 0.383 & 1232.640$\pm$413.737 & 2.403$\pm$0.523 \\
{} & 0.383 $-$ 1.791 & 645.221$\pm$33.499 & 3.783$\pm$0.300 \\
{} & 1.791 $-$ 4.287 & 408.177$\pm$12.684 & 2.176$\pm$0.079 \\
{} & 4.287 $-$ 6.015 & 286.788$\pm$16.786 & 1.273$\pm$0.066 \\
{} & 6.015 $-$ 8.511 & 214.702$\pm$14.967 & 1.071$\pm$0.068 \\
{} & 8.511 $-$ 11.263 & 192.329$\pm$22.508 & 0.873$\pm$0.075 \\
{} & 11.263 $-$ 18.495 & 178.013$\pm$35.499 & 0.780$\pm$0.093 \\
\hline
090809978 & -1.727 $-$ 1.023 & 382.599$\pm$206.356 & 1.709$\pm$0.424 \\
{} & 1.023 $-$ 1.983 & 255.765$\pm$50.279 & 2.427$\pm$0.282 \\
{} & 1.983 $-$ 4.863 & 209.860$\pm$15.778 & 1.502$\pm$0.049 \\
{} & 4.863 $-$ 8.127 & 131.351$\pm$13.241 & 0.831$\pm$0.034 \\
{} & 8.127 $-$ 11.263 & 67.924$\pm$13.042 & 0.512$\pm$0.045 \\
{} & 11.263 $-$ 20.159 & 40.761$\pm$29.170 & 0.383$\pm$0.105 \\
\hline
100612726 & -0.063 $-$ 1.087 & 181.979$\pm$56.287 & 1.221$\pm$0.153 \\
{} & 1.087 $-$ 2.943 & 132.842$\pm$9.858 & 1.309$\pm$0.070 \\
{} & 2.943 $-$ 5.247 & 105.570$\pm$4.407 & 0.857$\pm$0.029 \\
{} & 5.247 $-$ 6.143 & 85.145$\pm$8.302 & 0.568$\pm$0.044 \\
{} & 6.143 $-$ 7.487 & 54.283$\pm$8.528 & 0.381$\pm$0.038 \\
{} & 7.487 $-$ 9.791 & 61.054$\pm$39.131 & 0.341$\pm$0.050 \\
\hline
110817191 & -0.063 $-$ 0.447 & 395.760$\pm$118.835 & 2.528$\pm$0.644 \\
{} & 0.447 $-$ 0.831 & 316.608$\pm$48.255 & 3.065$\pm$0.459 \\
{} & 0.831 $-$ 1.919 & 193.306$\pm$12.979 & 1.872$\pm$0.105 \\
{} & 1.919 $-$ 2.943 & 124.718$\pm$11.275 & 1.235$\pm$0.080 \\
{} & 2.943 $-$ 4.415 & 57.360$\pm$9.035 & 0.622$\pm$0.052 \\
{} & 4.415 $-$ 7.551 & 29.179$\pm$11.825 & 0.430$\pm$0.081 \\
\enddata
\end{deluxetable}


\clearpage

\begin{figure*}[htbp]
\figurenum{4}
\centering
\gridline{\fig{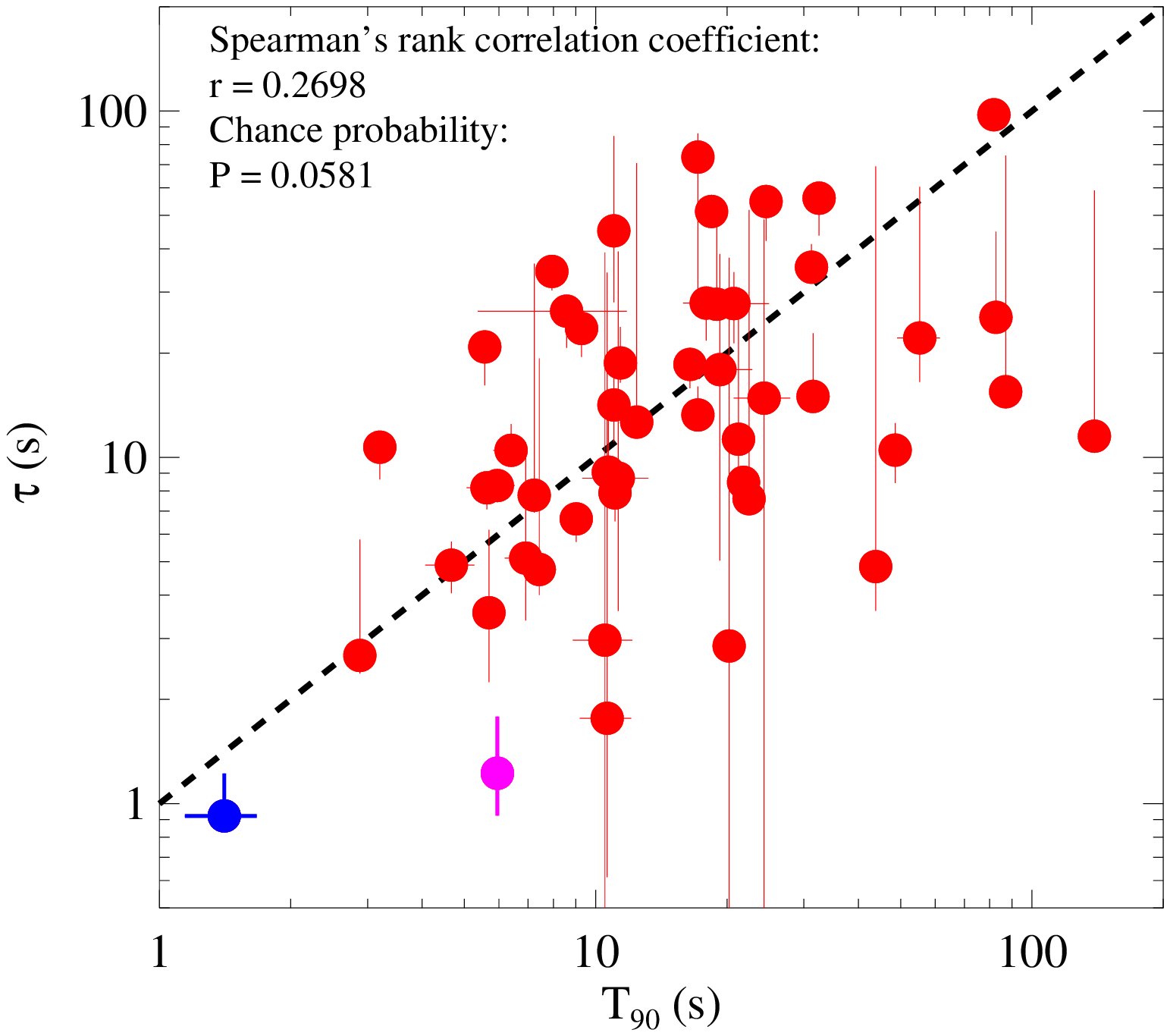}{0.45\textwidth}{(a)}
		\fig{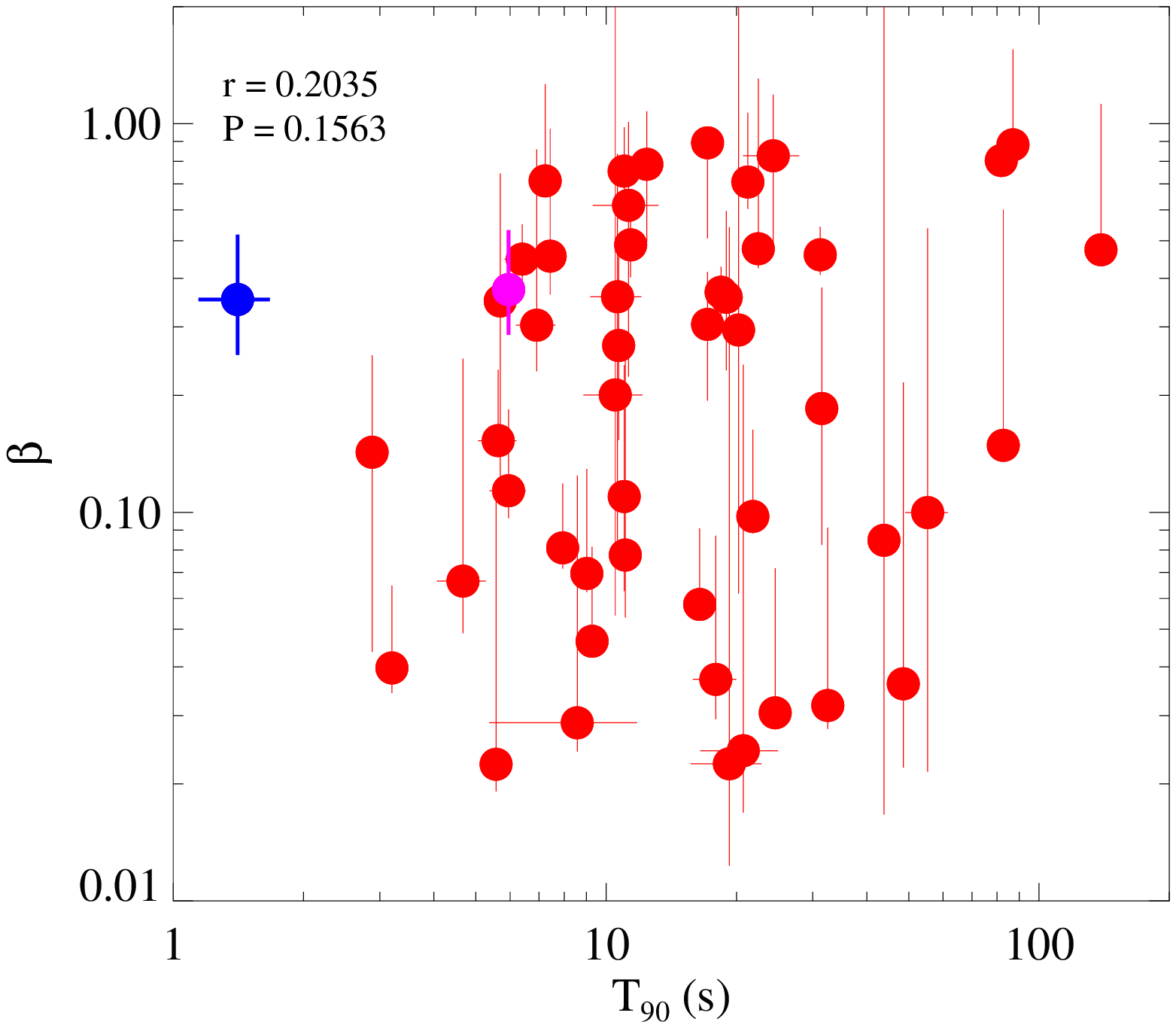}{0.45\textwidth}{(b)}
             }
\gridline{\fig{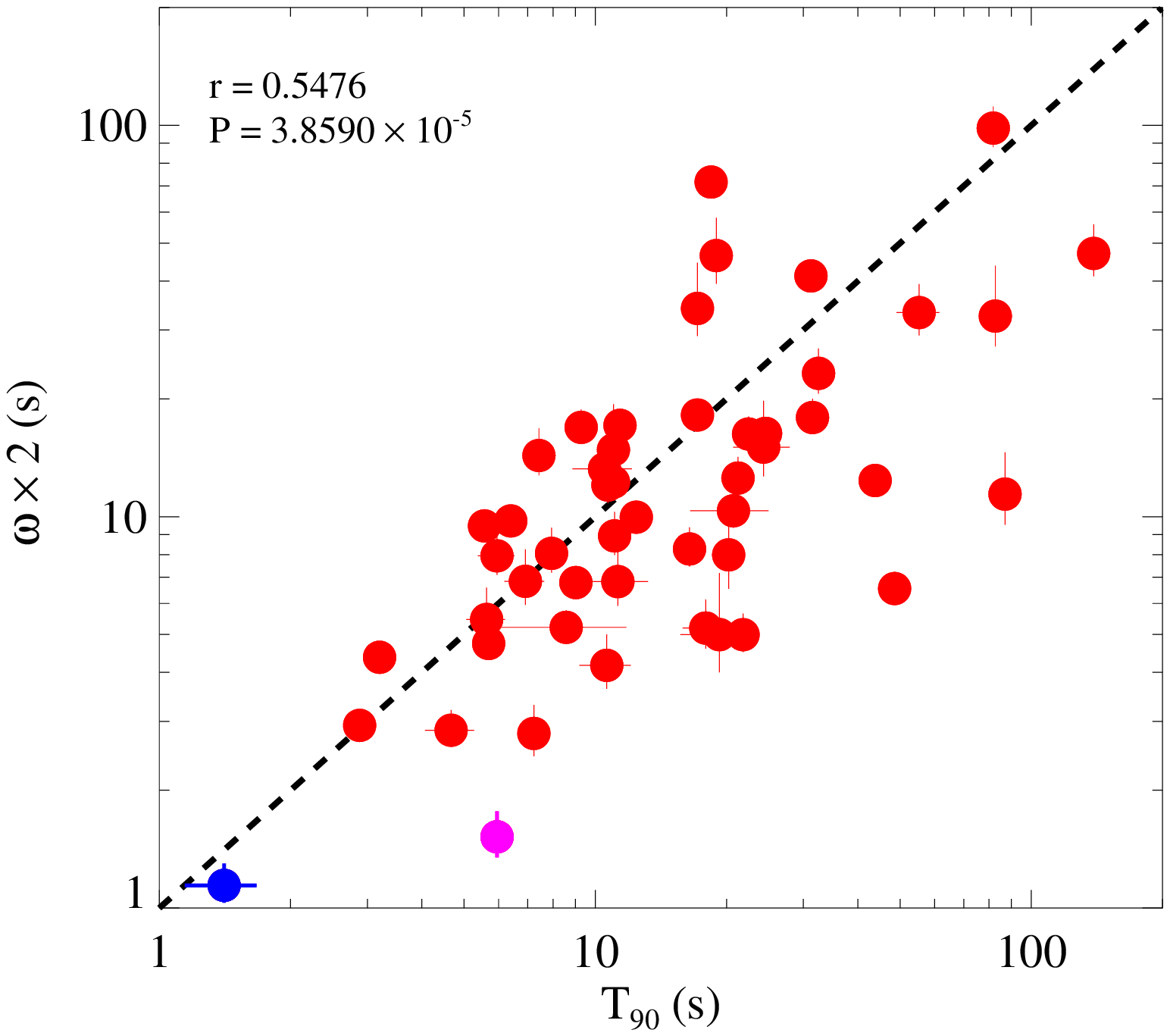}{0.45\textwidth}{(c)}
		\fig{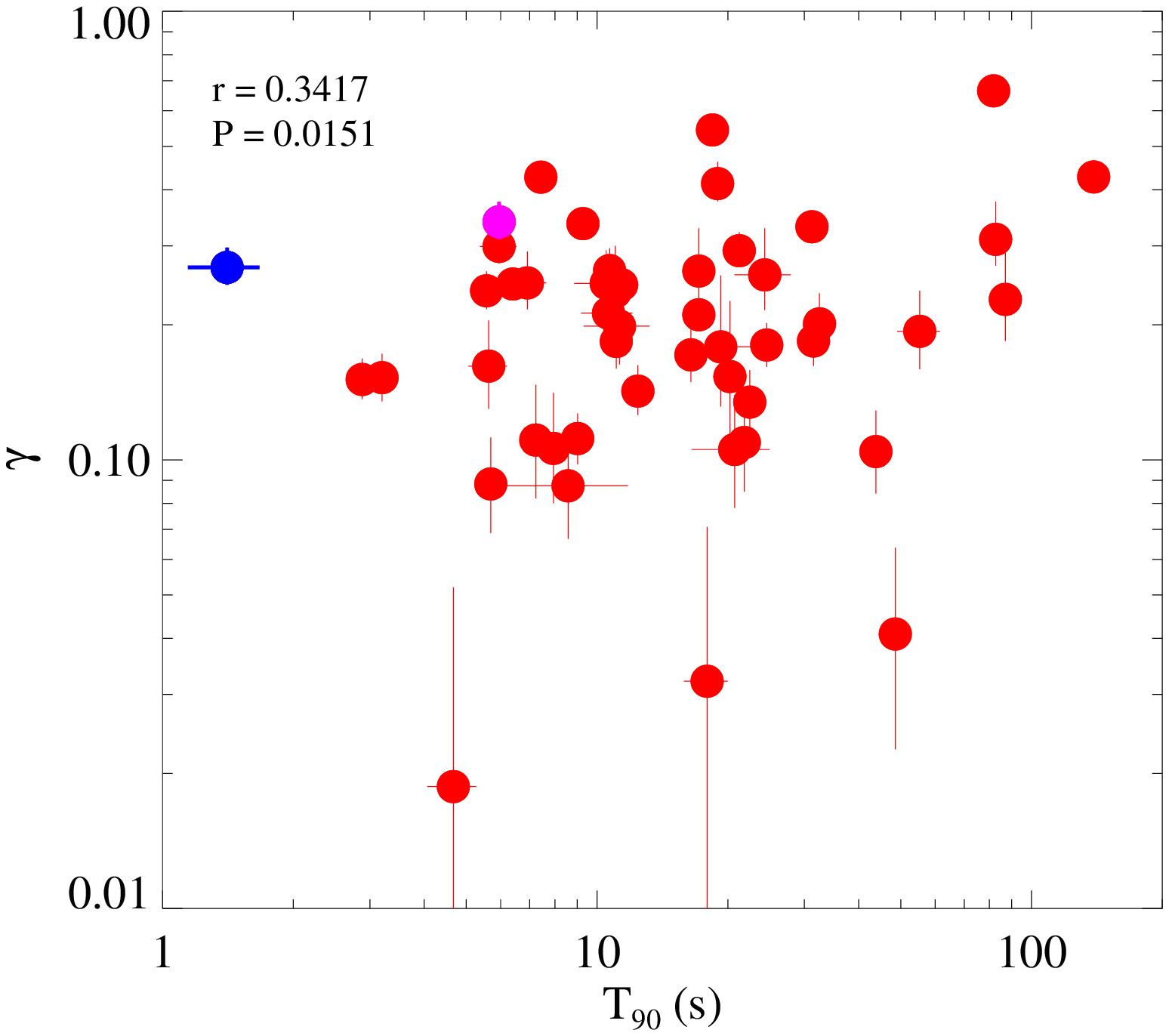}{0.45\textwidth}{(d)}
		}
\hfill
\caption{Scatter plots of $T_{90}$, $\tau$, $\omega$, $\beta$ and $\gamma$ for the 50 GRBs in our sample. The only short GRB 140209313 ($T_{90}=1.41$~s) is in blue points. Another special GRB 091010113 ($T_{90}=5.95$~s, but with an apparent pulse width less than 2~s) is in purple points. The rest are in red points. The dashed lines in panel (a), (c) and (e) indicate the line of equality. The Spearman's rank correlation coefficient $r$ and the corresponding chance probability $P$ are also shown in the upper left corner of each plot. \label{fig:correlation}}
\end{figure*}
\clearpage
\addtocounter{figure}{-1}
\begin{figure*}[htbp]
\figurenum{4}
\centering
\gridline{ \fig{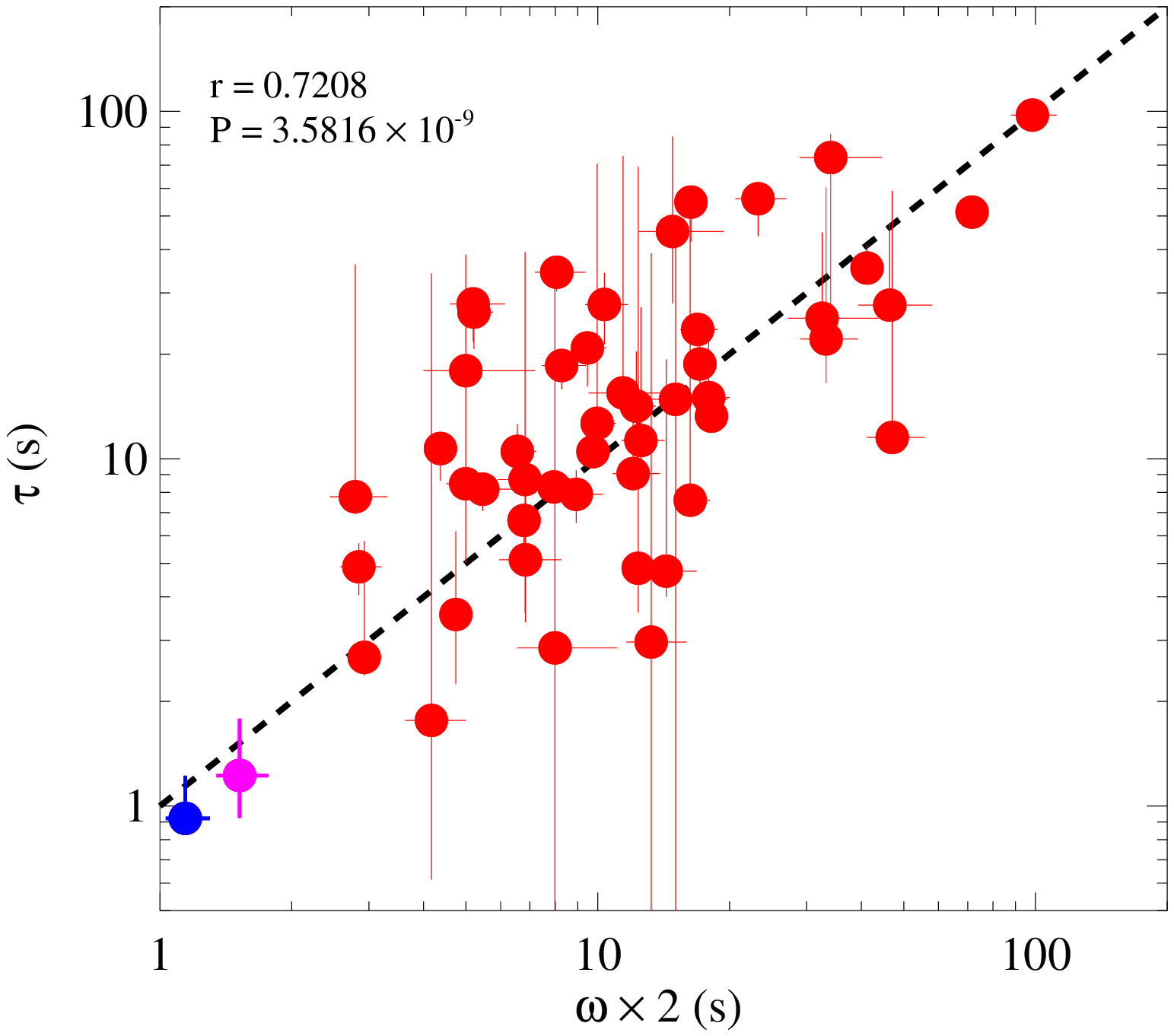}{0.45\textwidth}{(e)}
     \fig{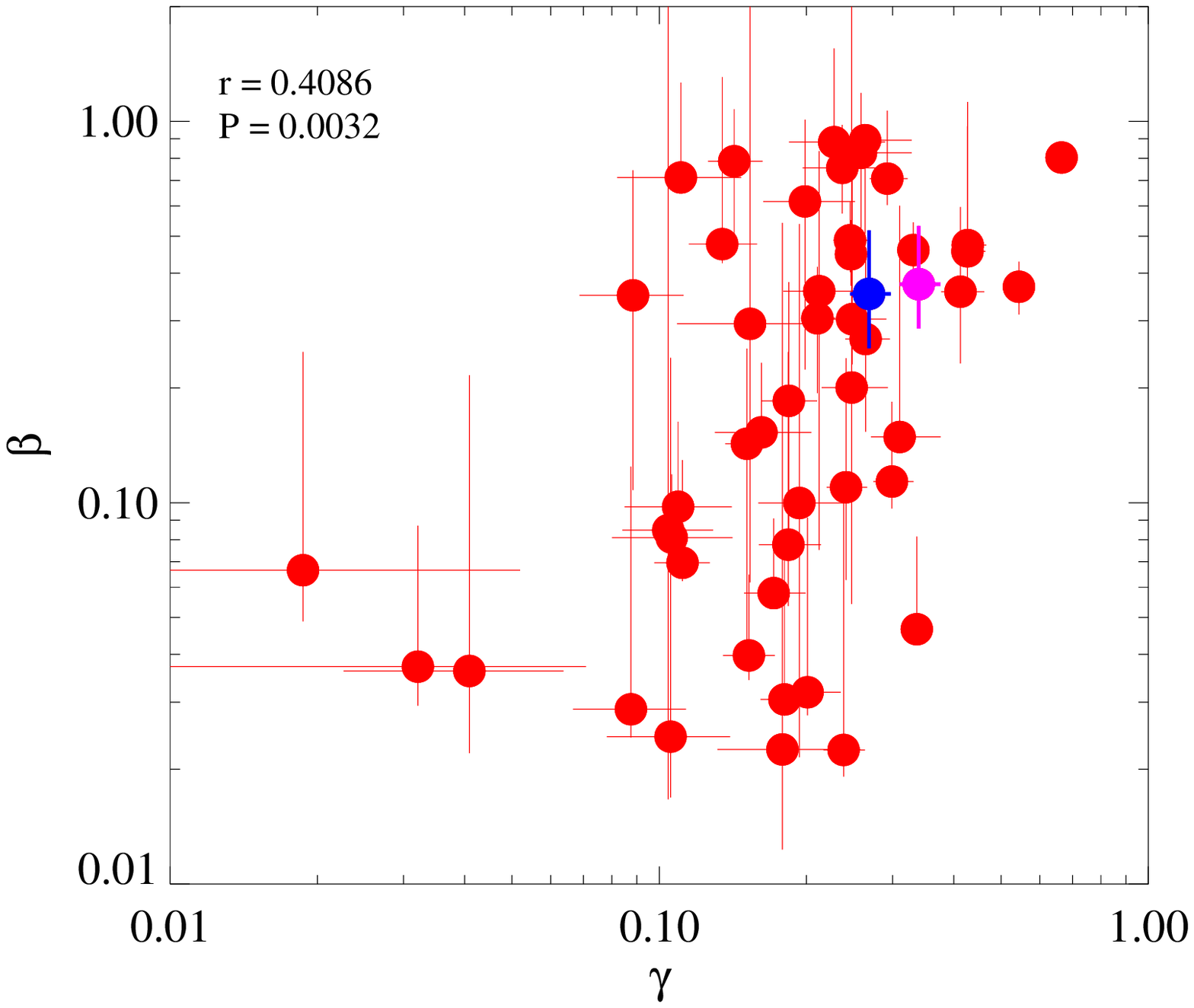}{0.45\textwidth}{(f)}
  			}   
\gridline{	\fig{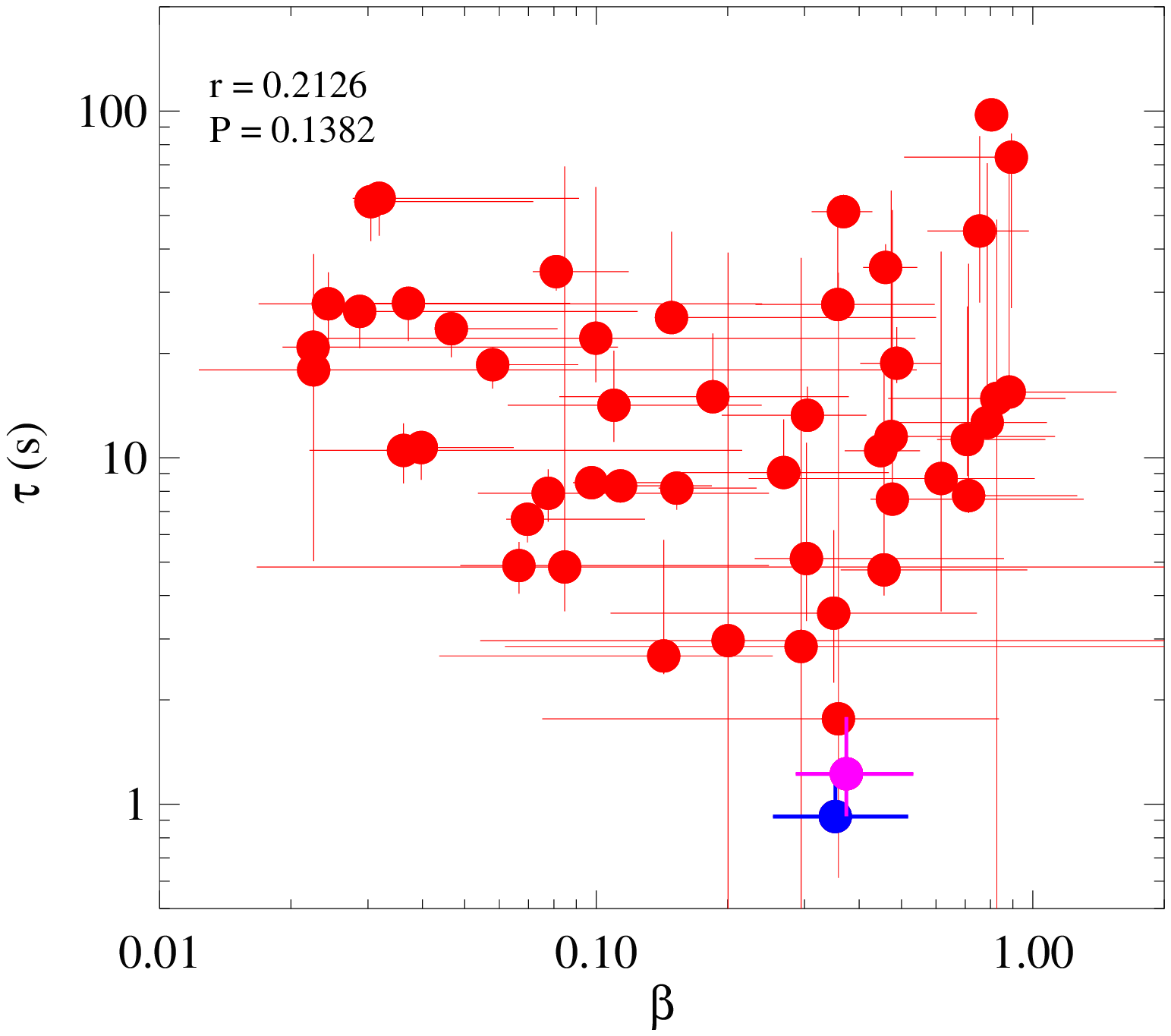}{0.45\textwidth}{(g)}
			\fig{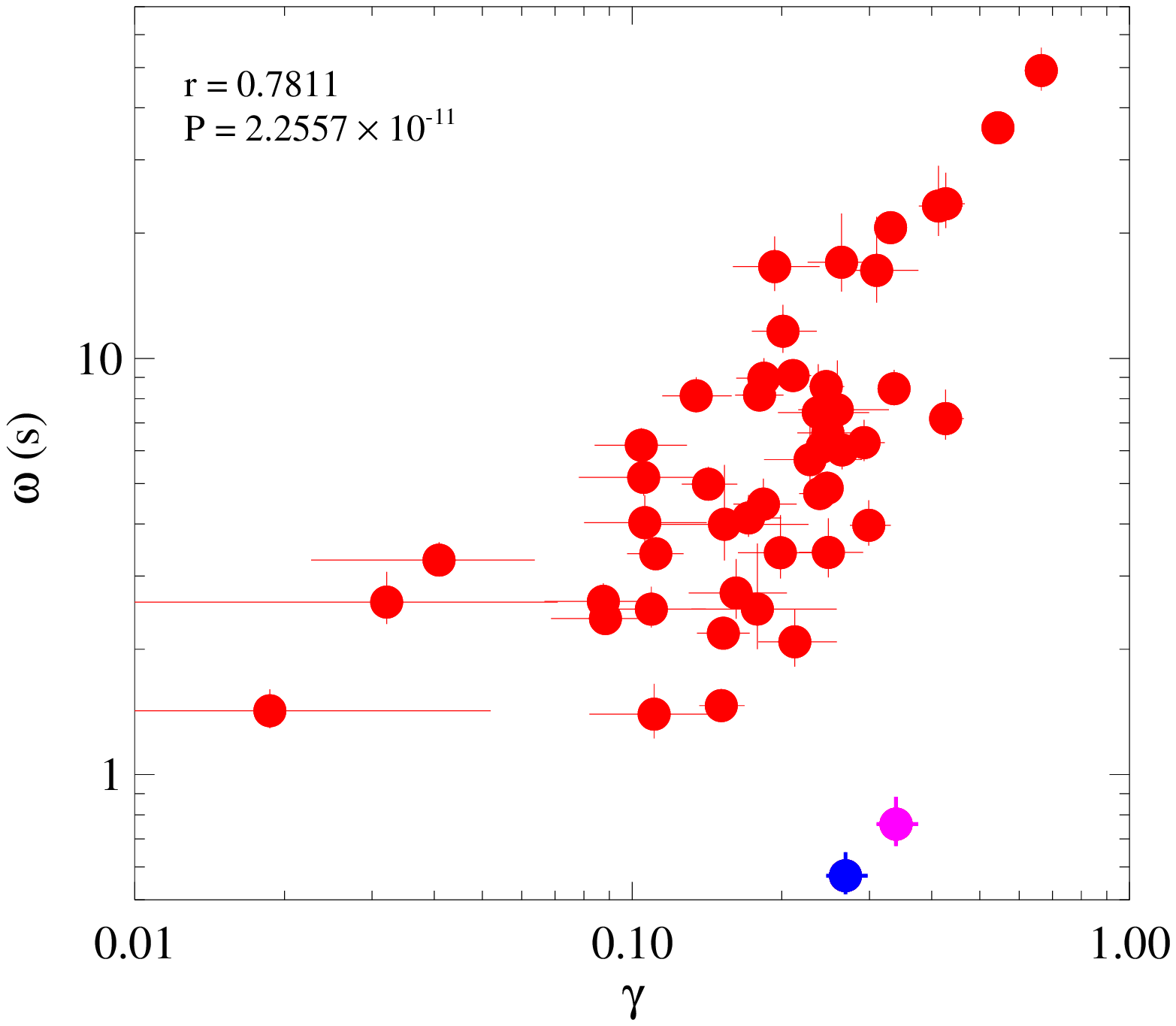}{0.45\textwidth}{(h)}
				}
\hfill
\caption{(Continued.)}
\end{figure*}
\clearpage
\addtocounter{figure}{-1}

\begin{figure}
\figurenum{5}
\gridline{\fig{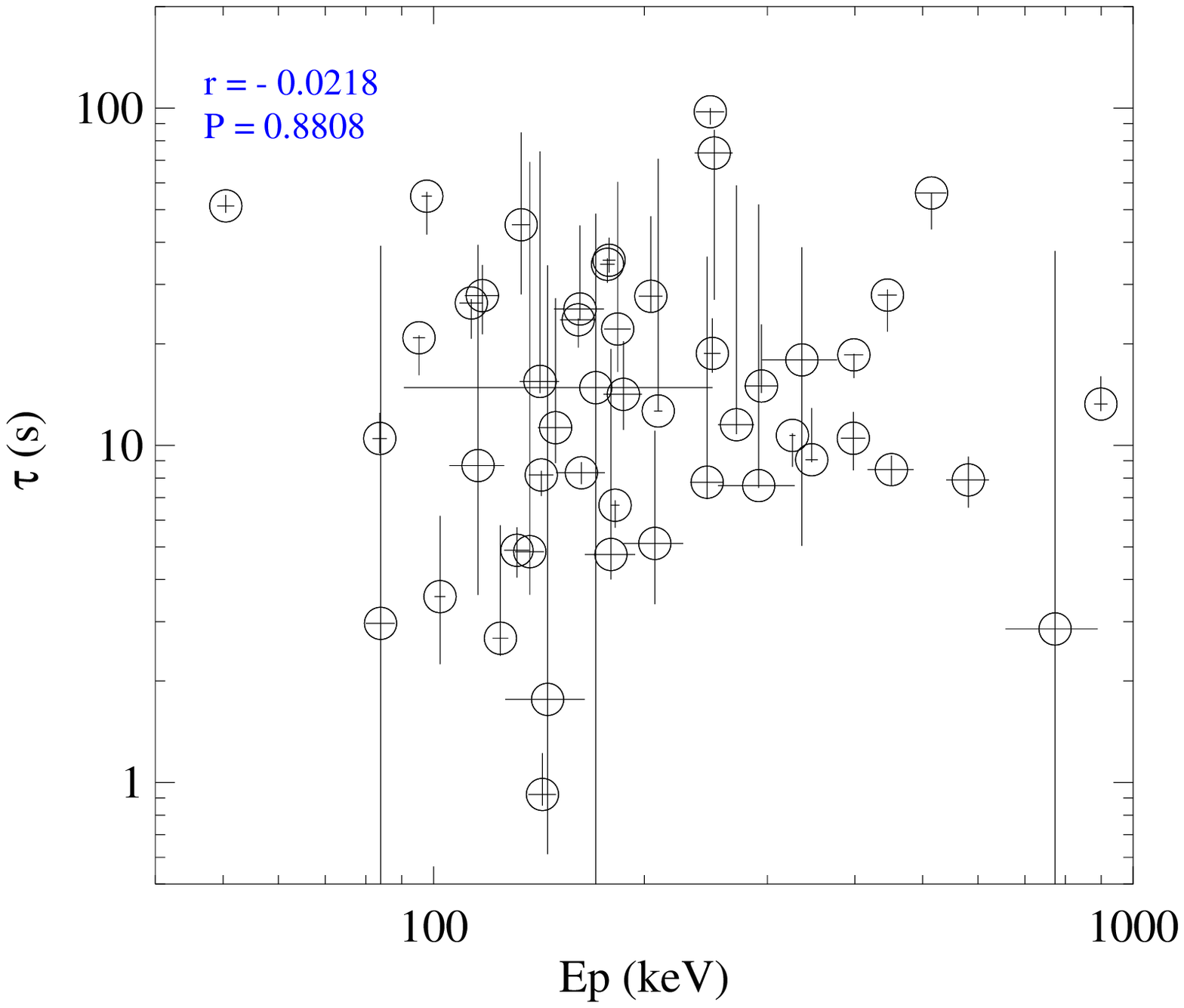}{0.45\textwidth}{(a)}
			\fig{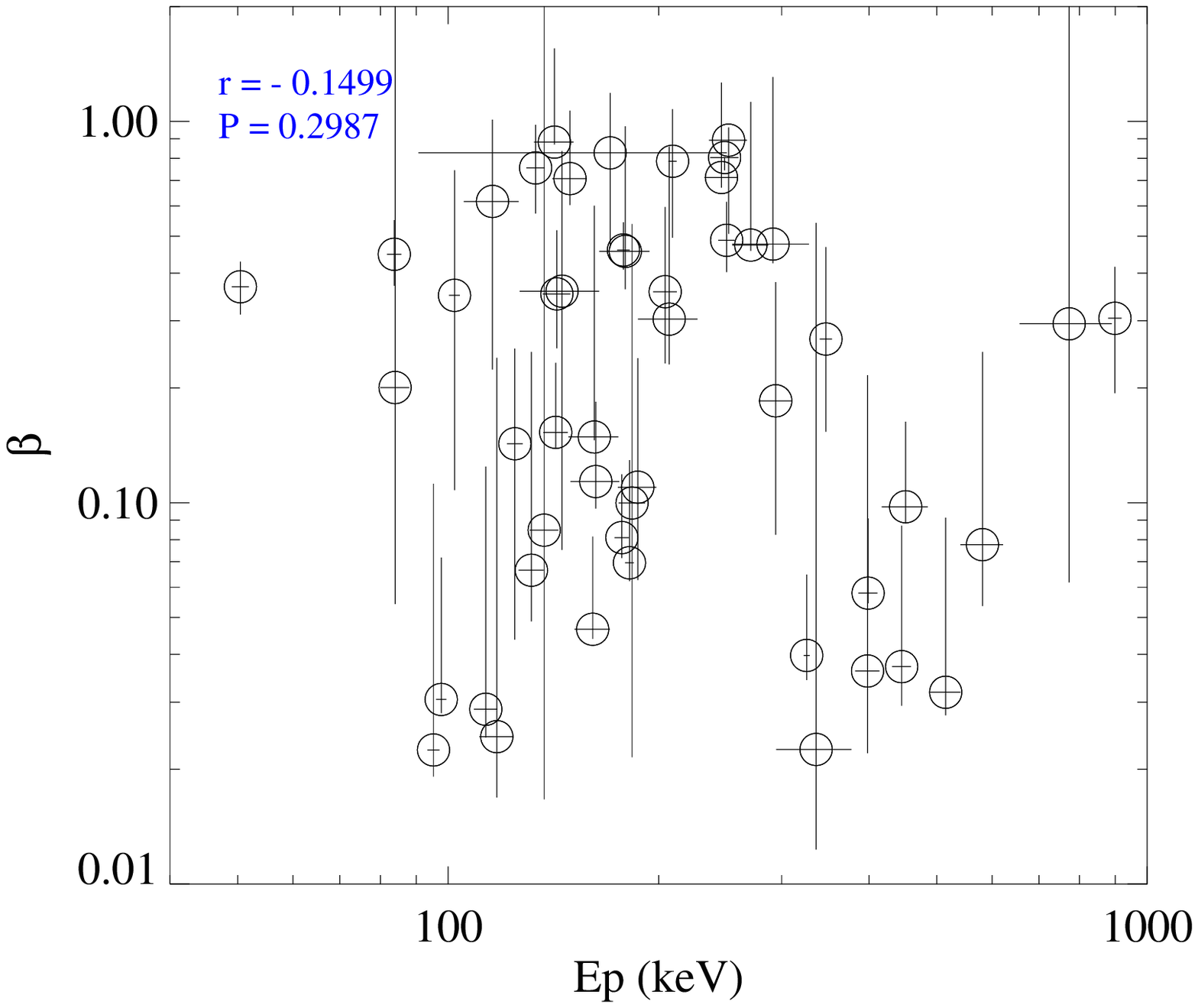}{0.45\textwidth}{(b)}
        }
\gridline{\fig{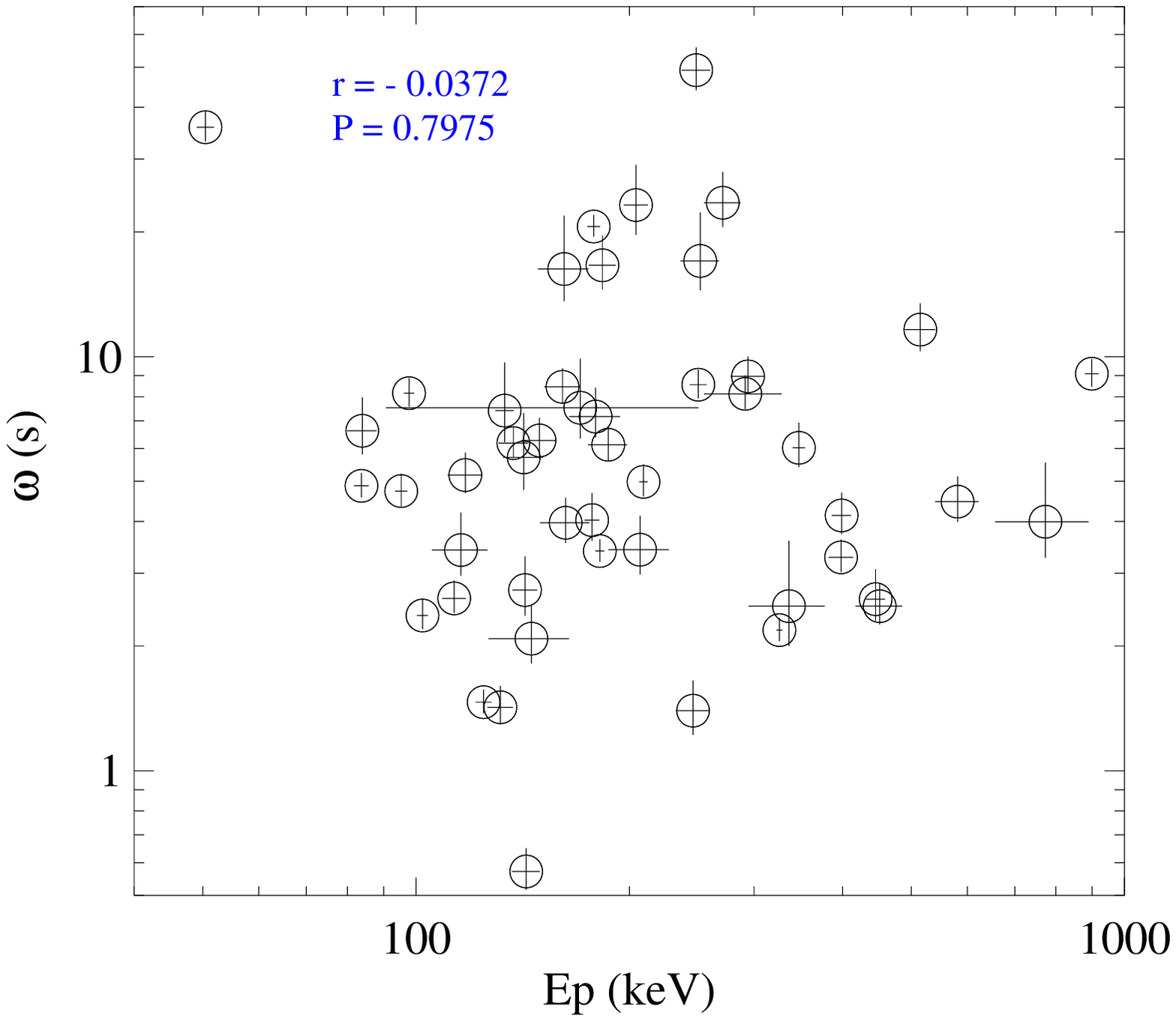}{0.45\textwidth}{(c)}
     \fig{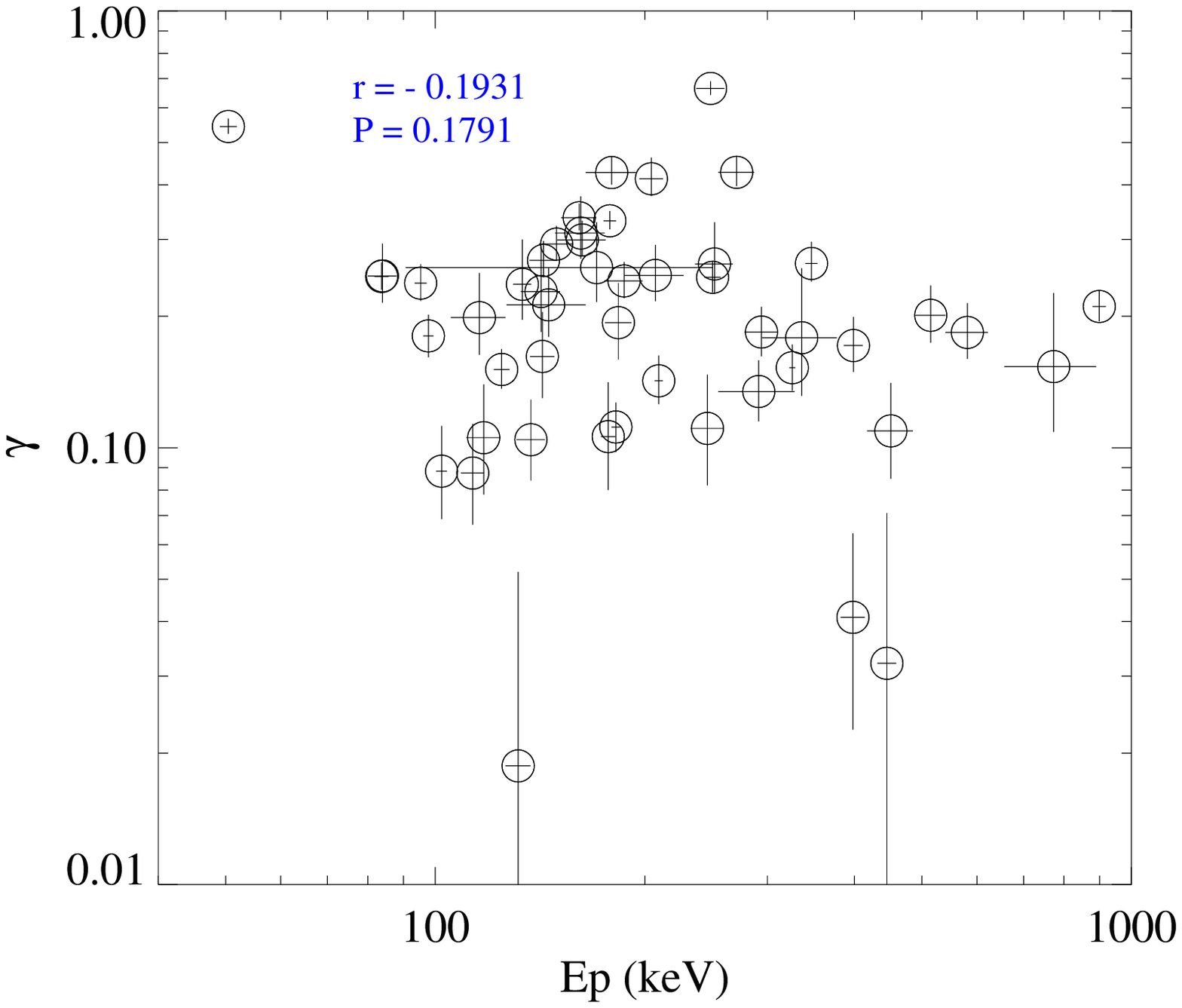}{0.45\textwidth}{(d)}
  			}   
\caption{Scatter plots of $\tau$, $\beta$, $\omega$ and $\gamma$ versus $E_{\rm p}$. The Spearman's rank correlation coefficient $r$ and the corresponding chance probability $P$ are also shown in the upper left corner of each plot.\label{fig:noncorrelation}}
\end{figure}


\begin{figure}
\figurenum{6}
\centering
\gridline{\fig{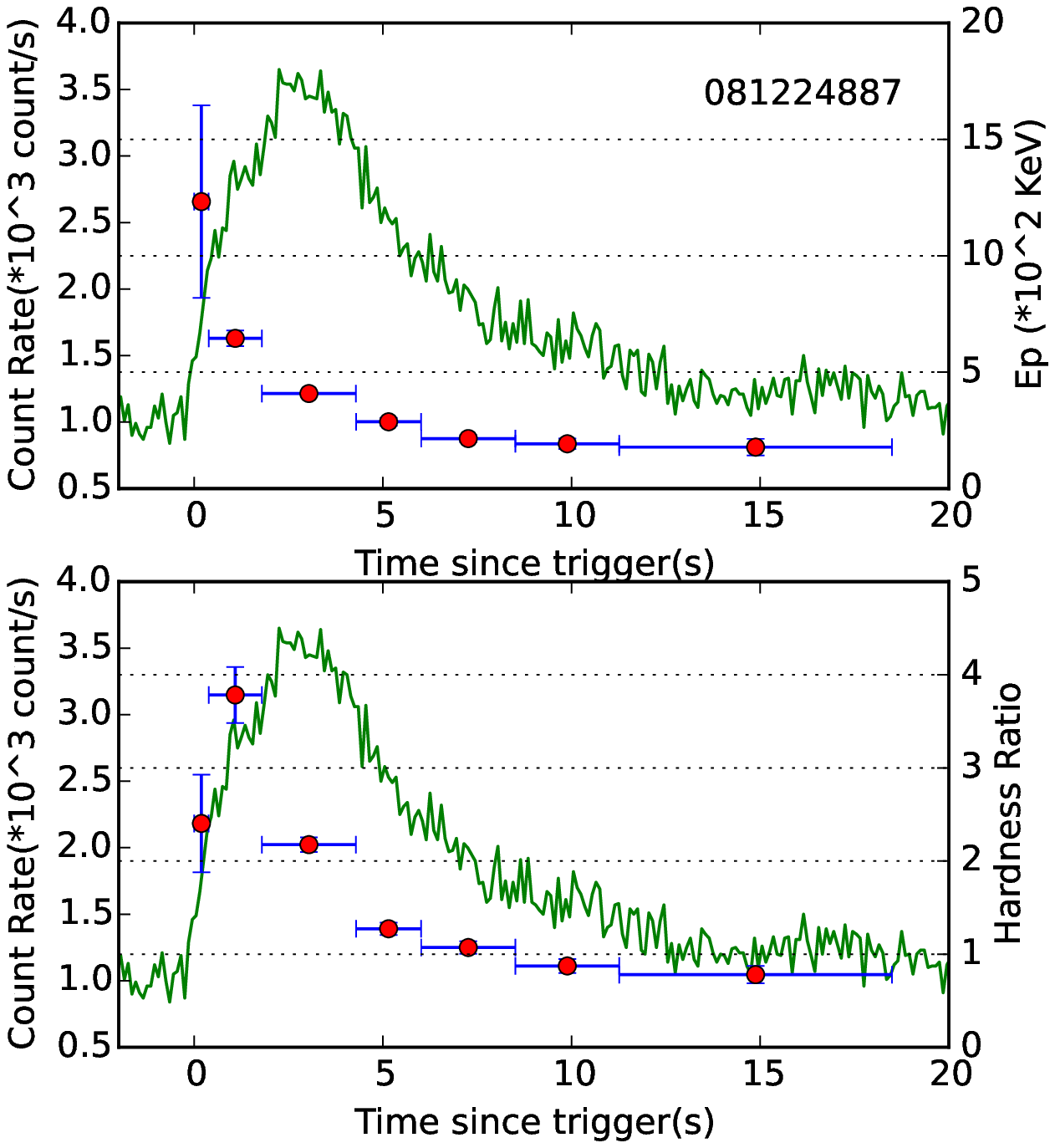}{0.35\textwidth}{(a) 081224887}
			\fig{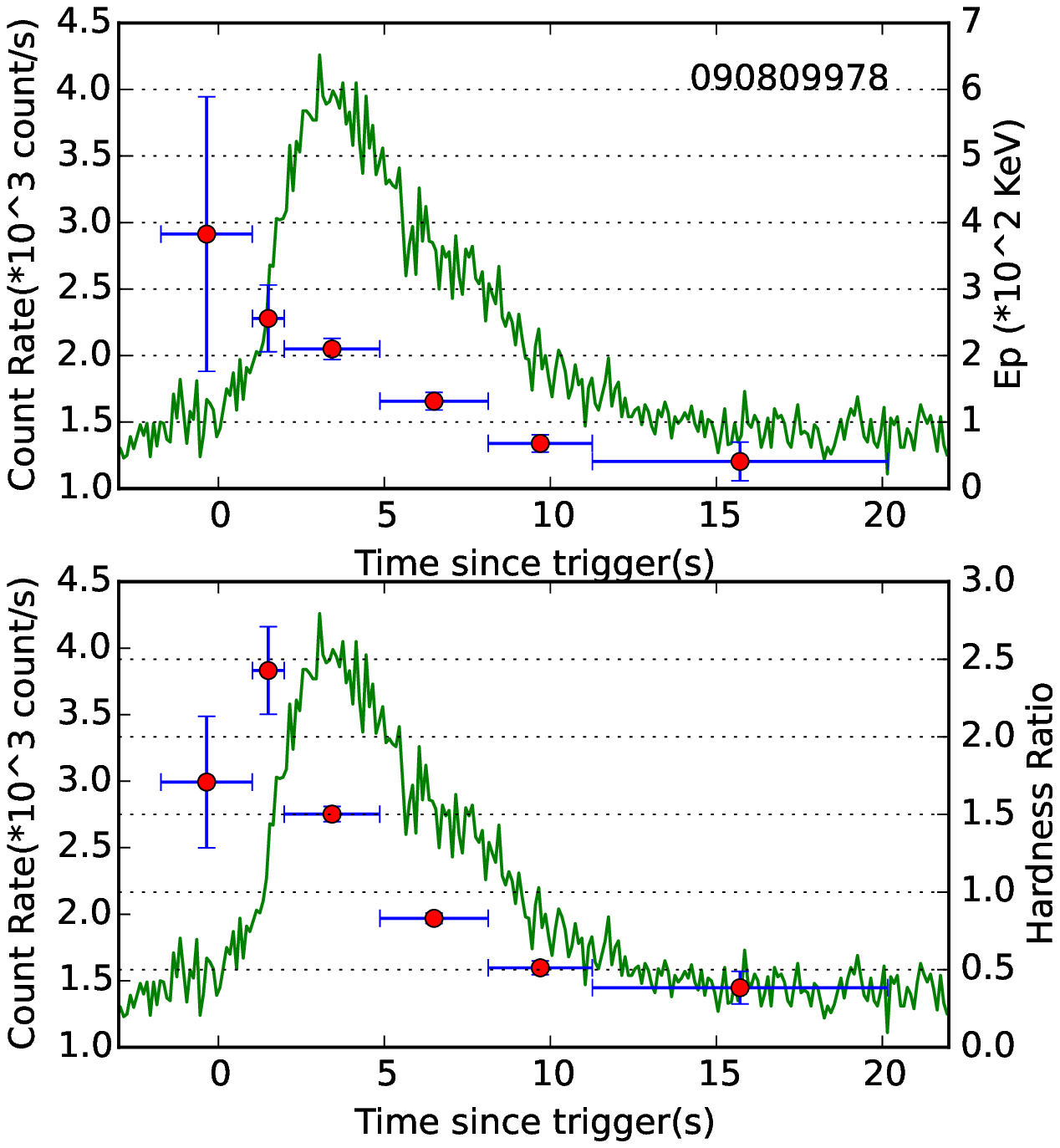}{0.35\textwidth}{(b) 090809978} }
\gridline{\fig{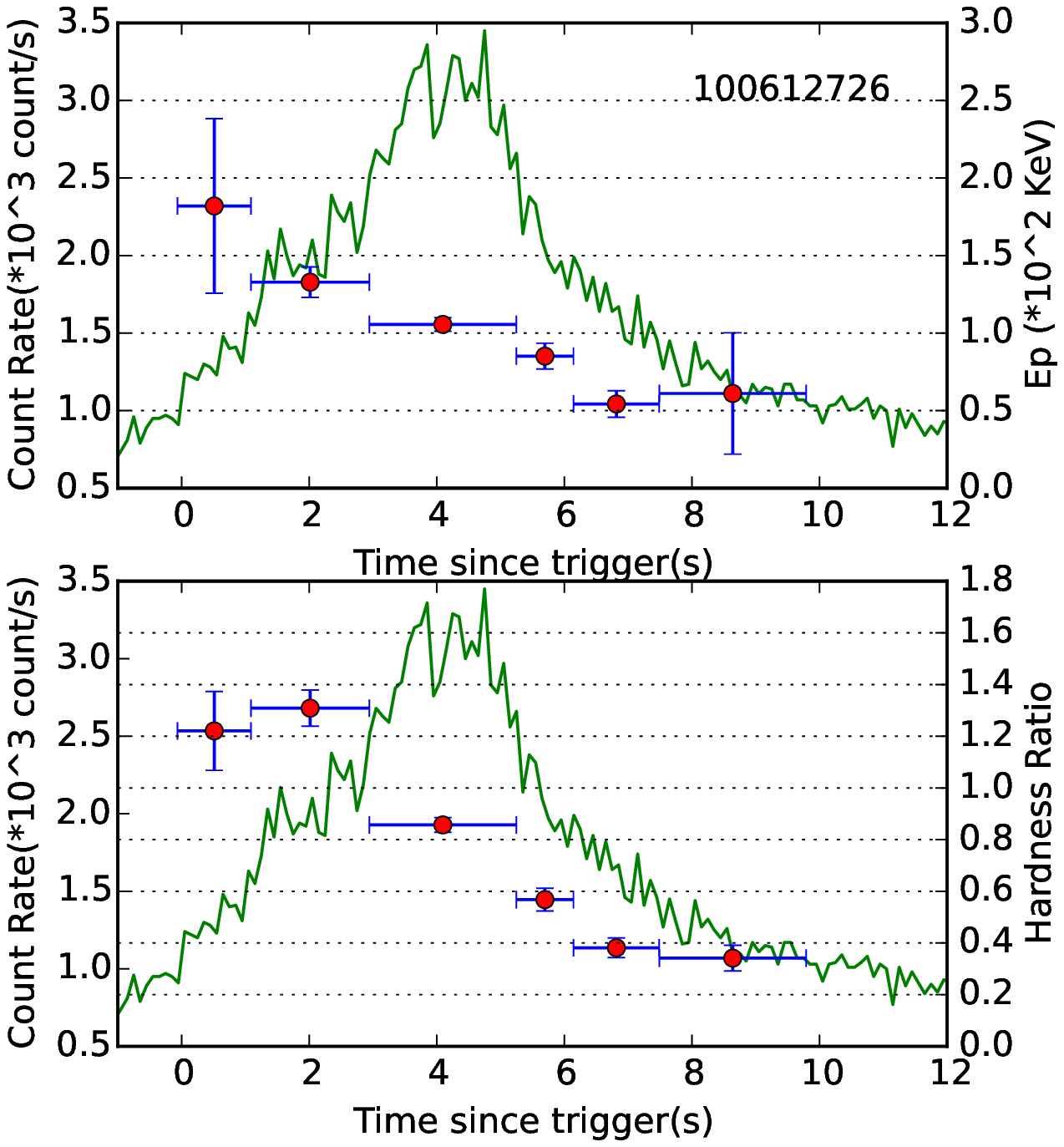}{0.35\textwidth}{(c) 100612726}
     \fig{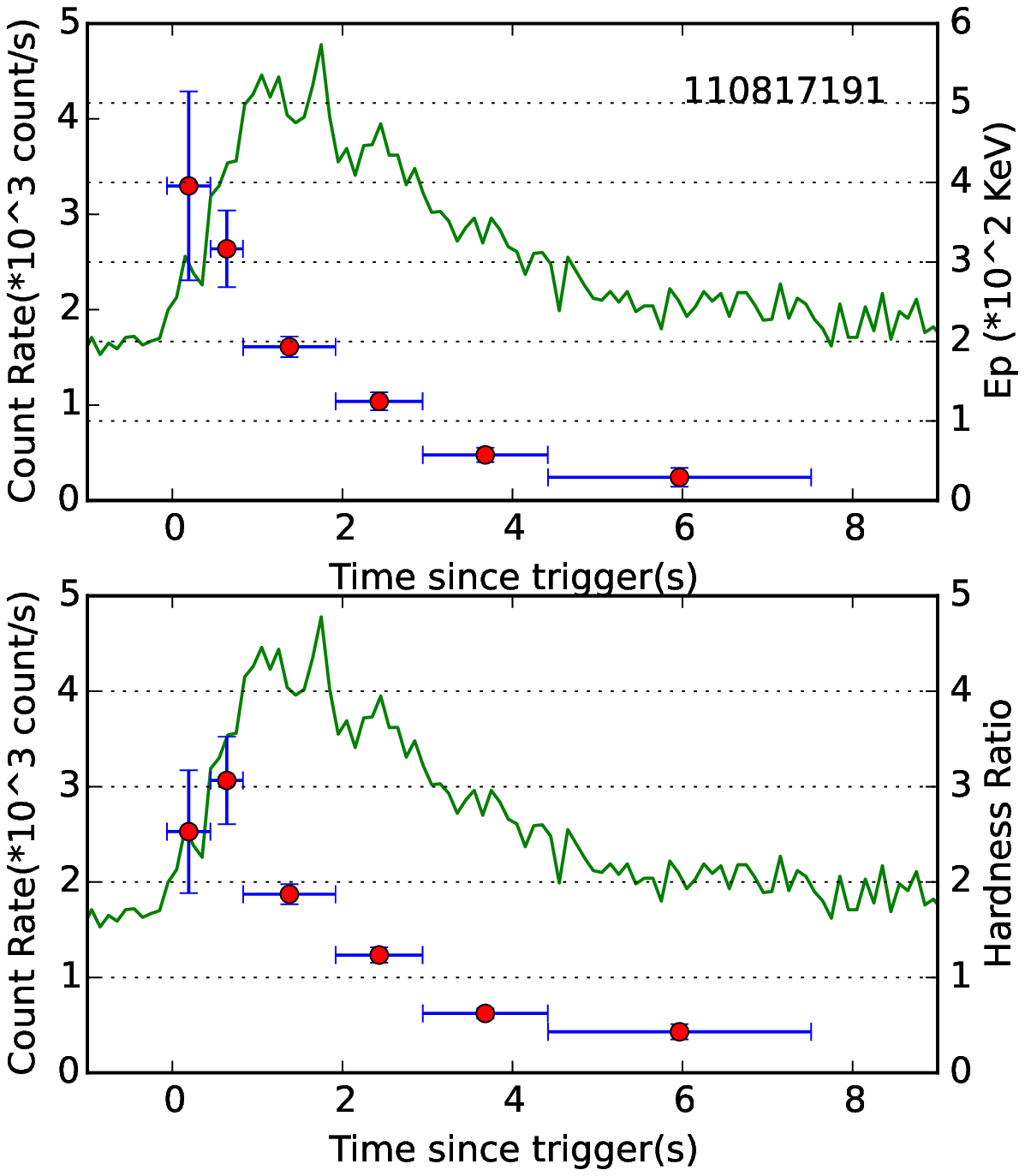}{0.35\textwidth}{(d) 110817191}  } 
\caption{Comparison of the evolution behaviors of the peak energy $E_{\rm p}$ and the hardness ratio for four ``Hard-To-Soft'' (HTS) bursts previously proposed. The upper plots of each panel show the evolution of $E_{\rm p}$ in red points compared with the light curves in green curve. The lower plots of each panel show the evolution of the hardness ratio in red points compared with the light curves in green curve. \label{fig:hardness}}
\end{figure}


\begin{figure}
\figurenum{7}
\fig{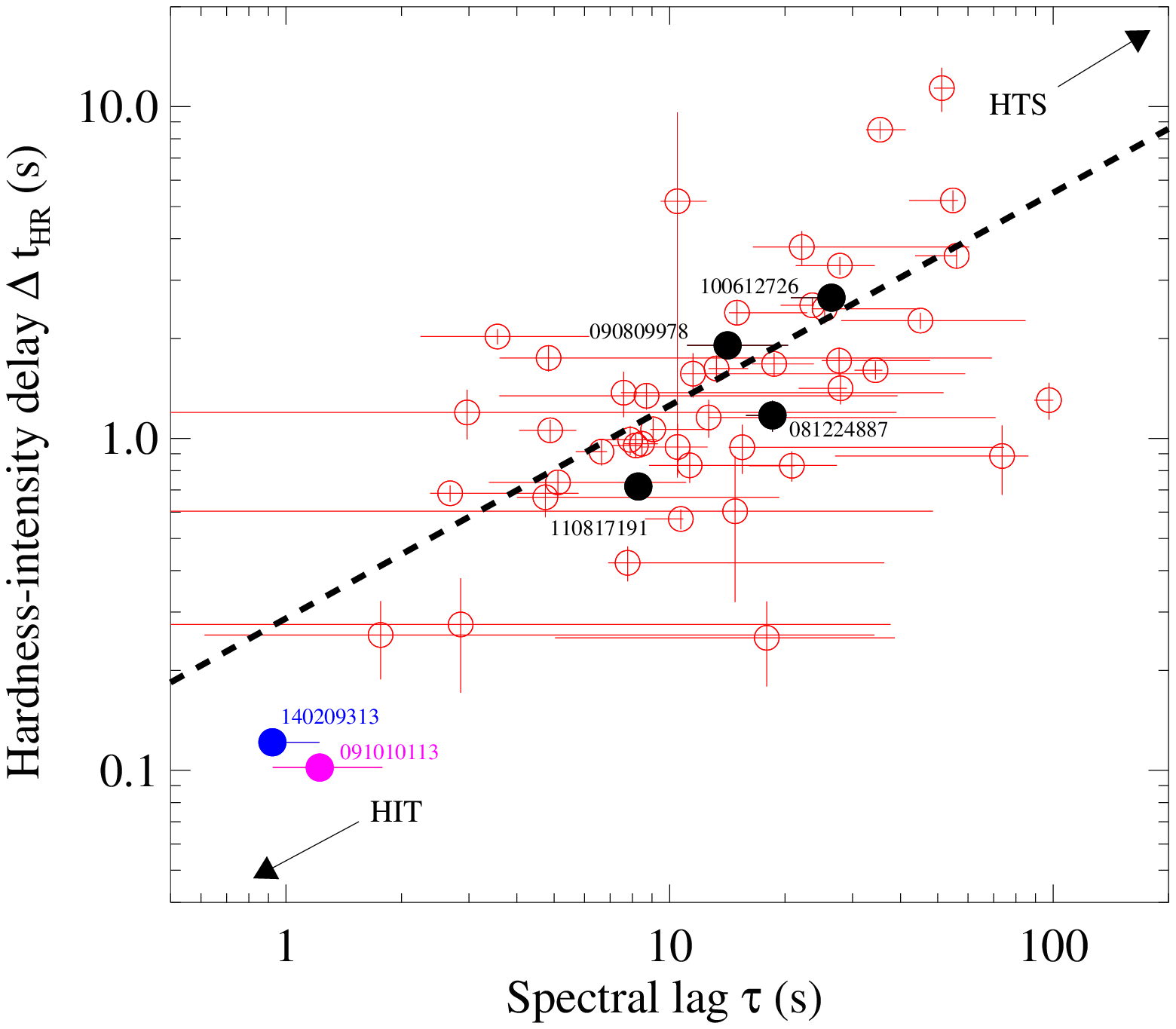}{0.8\textwidth}{}
\caption{Correlation between the hardness-intensity time delay $\Delta t_{\rm HR}$ and the spectral lag $\tau$. The best fit in dashed line is described by  Equation~(\ref{eq:HITlagcor}). The bursts in the lower left region have a negligible hardness-intensity delay, and therefore would show the HIT behavior. The bursts in the upper right region have a long hardness-intensity delay, and therefore would show the HTS behavior. Most bursts form a continuous transition between these two behaviors. The short GRBs 091010113 (in purple filled cirle) and 140209313 (in blue filled cirle) mentioned in Figure~\ref{fig:correlation} are in the lower left corner.  The four bursts mentioned in Figure~\ref{fig:hardness} are shown in black filled circles. \label{fig:HITlag}}
\end{figure}

\begin{figure*}[htbp]
\figurenum{8}
\centering
\gridline{\fig{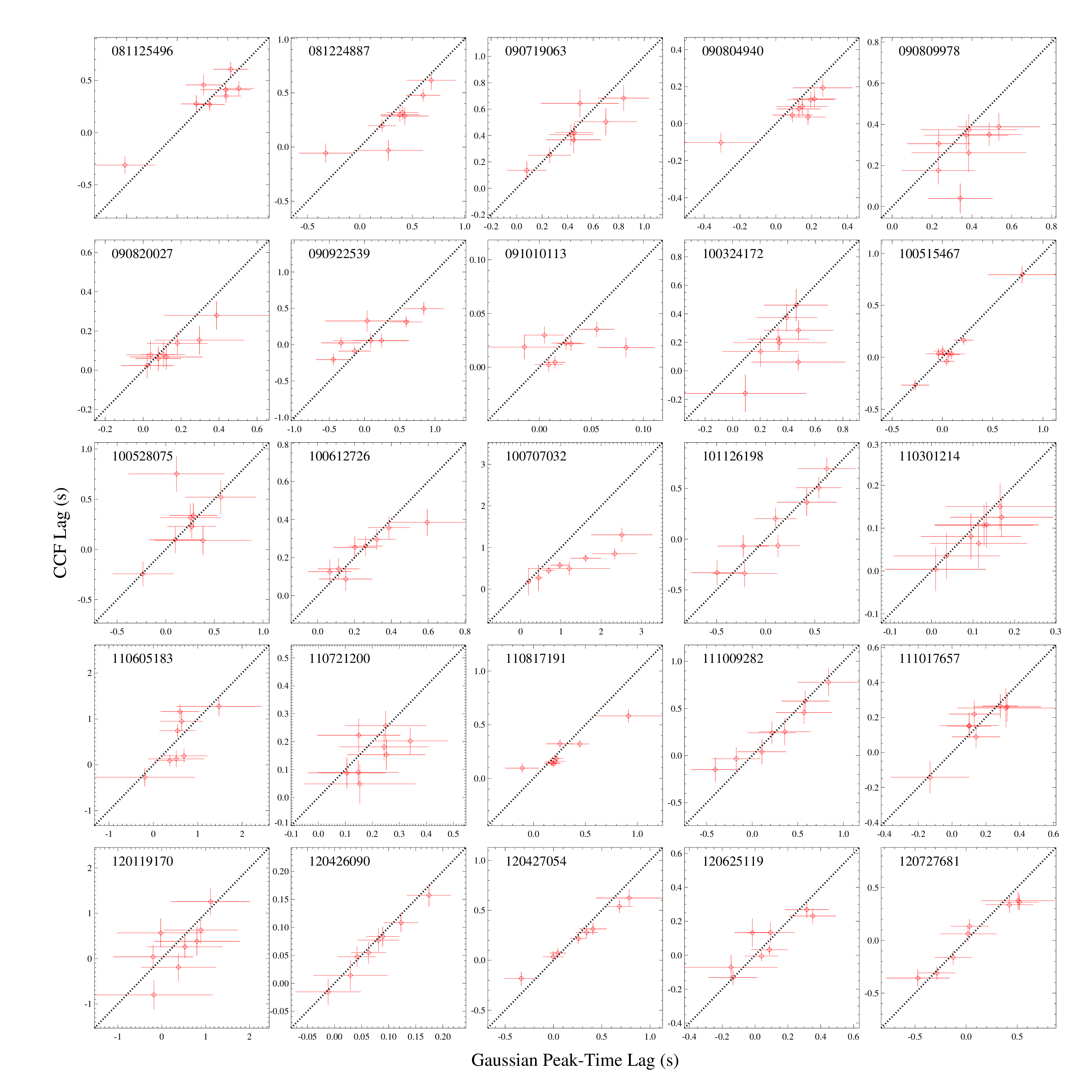}{0.95\textwidth}{}   }
\hfill
\caption{Comparision between the lags of two consecutive energy channels determined by CCF method and that determined by the difference between peak times of Gaussian profiles. The dashed lines represent the lines of equality where the results from both methods are consistent with each other.\label{fig:compare}}
\end{figure*}
\clearpage
\addtocounter{figure}{-1}
\begin{figure*}[htbp]
\figurenum{8}
\centering
\gridline{ \fig{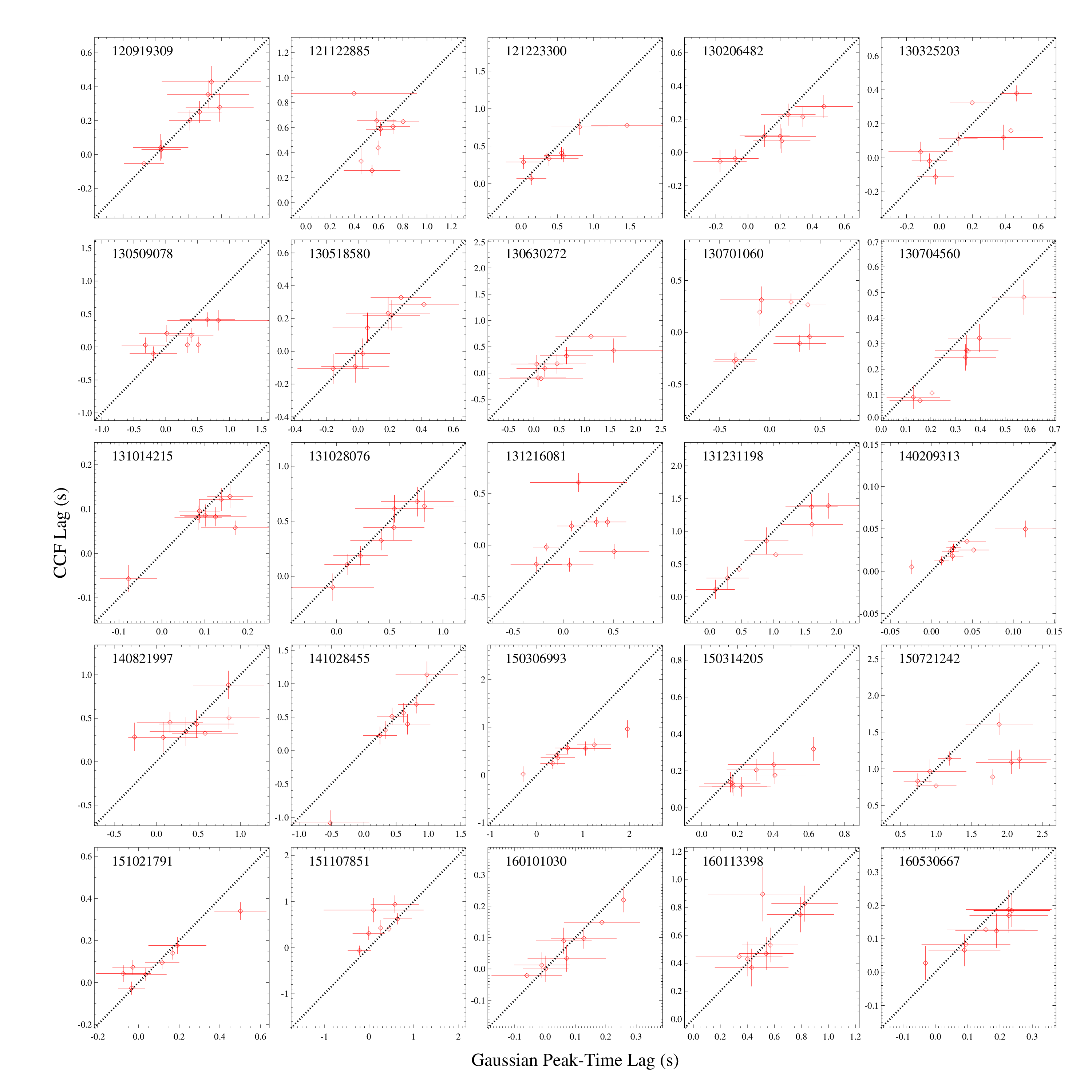}{0.95\textwidth}{} }
\hfill
\caption{(Continued.)}
\end{figure*}
\clearpage
\addtocounter{figure}{-1}

\end{document}